\documentclass{aa}

\usepackage{txfonts}
\usepackage{natbib}
\usepackage{amsmath,amsxtra}
\usepackage{booktabs}
\usepackage{graphicx}
\usepackage{multirow}
\usepackage[T1]{fontenc}

\usepackage{silence}
\usepackage[colorlinks, citecolor = blue, linkcolor = blue, urlcolor = blue]{hyperref}

\bibpunct[; ]{(}{)}{;}{a}{}{,} 

\begin{document}

\title{A low-luminosity type-1 QSO sample: IV. Molecular gas contents and conditions of star formation in three nearby Seyfert galaxies
}


\author{Lydia Moser \inst{1,3}
        \and Melanie Krips \inst{2}
        \and Gerold Busch \inst{1}
        \and Julia Scharwächter \inst{4}
        \and Sabine König \inst{2}
        \and Andreas Eckart \inst{1,3} 
        \and Semir Smaji\'c \inst{1,3}
        \and Macarena Garc\'ia-Marin \inst{1}
        \and Mónica Valencia-S. \inst{1}
        \and Sebastian Fischer \inst{5}
        \and Jens Dierkes \inst{6}     
}               
          

\institute{I. Physikalisches Institut, Universit\"at zu K\"oln, Z\"ulpicher Str. 77, 50937 K\"oln, Germany; \email{moser@ph1.uni-koeln.de}
        \and Institut de Radio Astronomie Millim\'etrique, 300 Rue de la Piscine, Domaine Universitaire, F-38406 Saint Martin d'H\`eres, France
        \and Max-Planck-Institut f\"ur Radioastronomie, Auf dem H\"ugel 69, 53121 Bonn, Germany
        \and LERMA (CNRS: UMR 8112), Observatoire de Paris, 61 Av. de l’Observatoire, 75014 Paris, France
        \and German Aerospace Center (DLR), K\"onigswinterer Str. 522-524, 53227 Bonn, Germany
        \and G\"ottingen eResearch Alliance, State and University Library G\"ottingen, Papendiek 14, 37073 G\"ottingen, Germany 
}

\date{Received 19 April 2015 / Accepted 15 September 2015 }

\abstract {We present a pilot study of $\sim$ 3$''$ resolution observations of low CO transitions with the Submillimeter Array 
in three nearby Seyfert galaxies, 
which are part of the low-luminosity quasi-stellar object (LLQSOs) sample consisting of 99 nearby (z =
0.06) type-1 active galactic nuclei (AGN)
taken from the Hamburg/ESO quasi-stellar object (QSO) survey.
Two sources
were observed in $^{12}$CO(2--1) and $^{13}$CO(2--1) and the third
in $^{12}$CO(3--2) and HCO$^+$(4--3).
None of the sources is detected in continuum emission.
More than 80\% of the $^{12}$CO detected molecular gas is concentrated within a diameter (FWHM) < 1.8 kpc. 
$^{13}$CO is tentatively detected,
while HCO$^+$ emission could not be detected. All three objects show indications of a kinematically decoupled central unresolved molecular gas component.
The molecular gas masses of the three galaxies are in the range $M_\textrm{mol} = (0.7 - 8.7) \times 10^9 M_\odot$. 
We give lower limits for the dynamical masses of $M_{\textrm{dyn}} > 1.5 \times 10^9 M_\odot$ and for the dust masses of $M_{\textrm{dust}} > 1.6 \times 10^6 M_{\sun}$. 
The $R_{21} = ^{12}$CO/$^{13}$CO(2--1) line luminosity ratios 
show Galactic
values of $R_{21} \sim 5-7$ in the outskirts and $R_{21} \gtrsim 20$ in the central region, similar to starbursts and (ultra)luminous infrared galaxies ((U)LIRGs; i.e. LIRGs and ULIRGs), implying higher temperatures and stronger turbulence. 
All three sources show indications of $^{12}$CO(2--1)/$^{12}$CO(1--0) ratios of $\sim 0.5$, suggesting a cold or diffuse gas phase. Strikingly, 
the $^{12}$CO(3--2)/(1--0) ratio of $ \sim 1$ also indicates a higher excited phase.
Since these galaxies have high infrared luminosities of $L_\textrm{IR} \geq 10^{11} L_\odot$ and seem to contain a circumnuclear starburst with minimum surface densities of gas and star formation rate (SFR) around $\Sigma_{mol}$ = 50 - 550 $M_\odot~$pc$^{-2}$ and $\Sigma_\textrm{SFR}$ = 1.1 - 3.1 $M_\odot~$kpc$^{-2}$yr$^{-1}$, we conclude that the interstellar medium in the centers of these LIRG Seyferts is strongly affected by violent star formation and better described by the ULIRG mass conversion factor.
}


\titlerunning{Molecular gas contents and conditions of star formation in three nearby Seyfert galaxies
}

\maketitle 



\section{Introduction}

The strong correlations between the mass of a supermassive black hole (SMBH) and the host galaxy properties, such as luminosity or central stellar velocity dispersion \citep[$i.e., M_\mathrm{BH}$--{$\sigma$} relation;][]{Gebhardt2000,MerFer2001,Tremaine2002,FerFord2005}, provide evidence that stellar bulges of galaxies and their SMBH are built up by mechanisms that are closely linked and suggests a coevolutional scenario \citep[e.g.,][]{Hopkins2008}.
A balance of nuclear fueling and feedback, in which the relative fractions of inflowing gas consumed by nuclear star formation and accretion onto a black hole are roughly constant, can explain the correlations \citep{Combes2009}.
However, the relevance of different feeding processes in decreasing angular momentum of the infalling material, the spatial scales they are working on, and their dependence on galaxy and active galactic nucleus (AGN) type is not fully understood \citep[i.e., as due to kpc-perturbations/mergers vs. secular evolution;][]{Hutchings1992,Sanders1988,Kormendy2004,Schawinski2011,Kocevski2012,GB2012,StorchiB2014}.

A detailed study of the gas content and its distribution in the host galaxies of AGN is essential for a deeper insight into the physical processes in the circumnuclear environment, e.g., the conditions for star formation and its properties \citep[e.g.,][]{Helfer2003,Iono2005,Krips2007,Bigiel2008,Leroy2008,Ford2013,Casasola2015}. 
The gas dynamics and kinematics of a galaxy population provide information about the mechanisms for nuclear fueling and, thus, help to improve fueling models. For example, central drops in gas/stellar velocity dispersion seem to go along with intense star formation activity and a more concentrated gas reservoir (r $\lesssim$ 500 pc), suggesting a dynamically cold (compared to the bulge) nuclear structure (e.g., a disk) to be an important fueling agent of nuclear activity \citep[e.g., star formation and AGN;][]{Falcon-Barroso2006,Hicks2013}.
Identifying links between the observations, at different redshifts and different evolutionary scenarios, is crucial for understanding the evolution of galaxies.


\begin{table*}[!htbp]
\caption{Properties of the three HE sources from literature}
\begin{tabular*}{\textwidth}{@{\extracolsep{\fill}} ccccccc}
\toprule
\multirow{2}{*}{Object} & \multirow{2}{*}{RA (J2000)\tablefootmark{1}} & \multirow{2}{*}{Dec. (J2000)\tablefootmark{1}} & \multirow{2}{*}{$z$} & \multirow{2}{*}{$\log (M_\textrm{BH}/M_{\sun})$} &  $L_\textrm{IR}$\tablefootmark{8} &  $M_\textrm{mol}$\tablefootmark{9}\\                 
 & & & & & [$10^{10}L_{\sun}$] & [$10^9 M_{\sun}$] \\
\midrule
%
%
%
%
HE 0433-1028 & 04$^\textrm{h}$36$^\textrm{m}$22.2$^\textrm{s}$ & $-$10\degr 22\arcmin 34\arcsec & 0.03555 $\pm$ 0.00001 \tablefootmark{2} & 6.9--8.3 \tablefootmark{5}&  27 &        9.0 \\                                                                                                          
HE 1029-1831 & 10$^\textrm{h}$31$^\textrm{m}$57.3$^\textrm{s}$ & $-$18\degr 46\arcmin 34\arcsec & 0.04026 $\pm$ 0.00009 \tablefootmark{3} & 6.7--7.4 \tablefootmark{6}&  25 &  2 - 12 \\                                                                                                          HE 1108-2813 & 11$^\textrm{h}$10$^\textrm{m}$48.0$^\textrm{s}$ & $-$28\degr 30\arcmin 04\arcsec & 0.02401 $\pm$ 0.00004 \tablefootmark{4} & 6.5--7.8 \tablefootmark{7}&   5 &   3.7 \\ 
\bottomrule
\end{tabular*}
\tablefoot{
\tablefoottext{1}{Values taken from NASA/IPAC Extragalactic Database (NED);}
\tablefoottext{2}{ \citet{Keel1996};}    
\tablefoottext{3}{ \citet{Kaldare2003};}
\tablefoottext{4}{ \citet{Theureau2005};}        
\tablefoottext{5}{ \citet[][X-ray variability]{Rao1992}, \citet[][optical lines]{Wang2007} and \citet[][optical lines]{Ryan2007};}
\tablefoottext{6}{ \citet[][optical/NIR lines and stellar velocity dispersion]{Busch2014};} 
\tablefoottext{7}{ \citet[][optical lines]{Wang2007};}
\tablefoottext{8}{ based on IRAS fluxes (NED) and formalism of \citet{Sanders1996} for the 8-1000 micron range;}
\tablefoottext{9}{ \citet{Bertram2007} and \citet{Krips2007}.
}
%
%
}
\label{lit_props}
\end{table*}

Cool molecular gas, the raw material for star formation, is traced best by low-J CO transitions. 
To characterize the gas excitation state and gas properties at high densities, higher CO transitions or high density tracers (e.g., HCN, HCO$^+$) and their isotopologues are necessary.  
These changes in interstellar medium (ISM) properties are already well traced by the $R_{10, 21} =$ $^{12}$CO/$^{13}$CO(1--0) or (2--1) and the $r_{31} =$  $^{12}$CO(3--2)/(1--0) luminosity ratio.
While these ratios are low in galactic disks with $R_{10, 21} \sim 5$ \citep[e.g.,][]{Solomon1979} and $r_{31} \sim 0.4 - 0.6$ \citep{Mauersberger1999,Israel2005}, the values increase toward the centers of galaxies and can get as high as $R_{10} > 20$ \citep[i.e., turbulent and/or hot gas state;][]{Devereux1994,Dumke2001,Muraoka2007,Papadopoulos2008,Iono2009,Papadopoulos2012a} and $r_{31} \gtrsim 1$ \citep[very warm, dense gas;][]{Huttemeister2000,Sakamoto2007,Aalto2007,Papadopoulos2008,Aalto2010,Costagliola2011} in 
(ultra)luminous infrared galaxies ((U)LIRGs; i.e. LIRGs and ULIRGs). 
In addition to (U)LIRGs, high values of $r_{31}$ (up to 5) are also found in centers of normal galaxies, starbursts, and AGN \citep[e.g.,][]{Devereux1994,Matsushita2004,Mao2010,Combes2013,GB2014}.
The contribution of diffuse gas, as well as of higher excitation phases to the brightness of line emission, becomes important in starburst and (U)LIRGs. These phases have a significant impact on the gas mass estimate and are signs of enhanced star formation activity \citep[e.g.,][]{Hinz2006,Papadopoulos2012b,Papadopoulos2012a,Bolatto2013,Kamenetzky2014,Shetty2014}.  

In this paper, we present a pilot study of CO emission in three galaxies (two in $^{12}$CO(2--1) and $^{13}$CO(2--1) and one in $^{12}$CO(3--2) and HCO$^+$(4--3) at a moderate angular resolution with the aim of probing the morphology, kinematics, and physical conditions of the ISM. 
The three galaxies are part of a large, representative sample of 99 nearby type-1 AGN from the Hamburg/ESO quasi-stellar object (QSO) survey \citep{Wisotzki2000} with a redshift cutoff of $z\leq 0.06$ \citep[][]{Bertram2007} so that near-infrared (NIR) diagnostic lines for the stellar and gaseous content are still accessible \citep[e.g.,][]{Gaffney1995,Fischer2006,Busch2014}.
Because of their $B_\mathrm{J}$ magnitudes around the traditional Seyfert/QSO demarcation \citep{Koehler1997}, these sources are also called low-luminosity quasi-stellar objects (LLQSOs) to emphasize their transient nature.

The goal of the LLQSO sample is to explore the signatures of internal or external triggers of the (circum-) nuclear activity (i.e., star formation and accretion onto the SMBH) down to sub-kiloparsec scales, their importance in LLQSOs, and with this, the link of this population to local 
low-luminosity active galactic nuclei (LLAGN)
and powerful QSOs at higher redshifts.

Observations of subsamples in $^{12}$CO(1--0), (2--1) and \ion{H}{i} \citep{Krips2007,Bertram2007,Konig2009}, in the NIR \citep{Fischer2006,Busch2014a,Busch2014}, and in the optical \citep[][]{Scharwaechter2011, Tremou2015} suggest that ongoing circumnuclear star formation plays an important, if not even dominant, role in the ISM characteristics and correlates with a large molecular gas reservoir. Judging by the activity level (e.g., $B_\mathrm{J}$ magnitude) and the FIR and CO luminosities, these sources seem to mark a transition population between local Seyfert/nonactive galaxies and higher-$z$ QSOs.

The three galaxies we discuss here belong to the most luminous galaxies in CO emission from our $^{12}$CO tested subsample \citep{Krips2007,Bertram2007}, which is why they were chosen for the follow-up observations.
Their $^{12}$CO(1--0) luminosities are larger than $L'_\mathrm{CO}= 0.9 \times 10^9 $ K km s$^{-1}$ pc$^2$, implying that they are very rich in molecular gas with masses larger than $M_{\mathrm{mol}}=3.7 \times 10^9 M_{\sun}$.
In Table \ref{lit_props} we list the molecular gas masses, including helium recalculated to standard cosmology with $H_0=70$, $\Omega_M=0.3$ and $\Lambda_0=0.7$. 
Their IRAS-based IR luminosities (8 - 1000$\mu$m) are located closely around the LIRG demarcation of $L_\mathrm{IR}=10^{11} L_{\sun}$. HE 0433-1028 and HE 1108-2813 have only been observed with single dish observations before \citep{Bertram2007}.
Their high $^{12}$CO luminosities make them ideal candidates for interferometric follow-up studies with the Submillimeter Array \citep[SMA,][]{Ho2004}.


\begin{table*}[!htbp]
\centering
\caption{Observational parameters}
\tabcolsep=0.11cm
\begin{tabular*}{\textwidth}{@{\extracolsep{\fill}} ccccccccccc}
\toprule
Object & \multicolumn{2}{c}{Phase tracking center} & Ant\tablefootmark{1} & $\nu_\textrm{CO rest}$ & $t_\textrm{total}$ &  $FOV$ & $\theta_\textrm{beam}$         & $rms_\textrm{cont}$   & $rms_{\textrm{5~km~s}^{-1}}$  & Date \\ 
         & RA (J2000) &  Dec. (J2000) &  & [GHz] & [h] &  [$''$] & [$''\times''$] & \multicolumn{2}{c}{[mJy beam$^{-1}$]}& \\ 
\midrule                                                                                                                                                                                
HE 0433-1028 & 04$^\textrm{h}$36$^\textrm{m}$22.20$^\textrm{s}$ & $-$10\degr 22\arcmin 32.9996\arcsec & 6 & 222.624 & 2.5 &   56 & 3.9 $\times$ 1.6 & 1.4 & 29 - 30\tablefootmark{2} & June 2008  \\  
HE 1029-1831 & 10$^\textrm{h}$31$^\textrm{m}$57.30$^\textrm{s}$ & $-$18\degr 46\arcmin 33.1965\arcsec & 7 & 332.432 & 2.3 &   37 & 2.5 $\times$ 1.5 & 3.7 & 60\tablefootmark{2} - 65 & April 2007 \\  
HE 1108-2813 & 11$^\textrm{h}$10$^\textrm{m}$48.00$^\textrm{s}$ & $-$28\degr 30\arcmin 02.9978\arcsec & 8 & 225.132 & 5.0 &   55 & 4.6 $\times$ 3.0 & 0.9 & 17 - 23\tablefootmark{2} & March 2008 \\  
\bottomrule
\end{tabular*}
\tablefoot{
\tablefoottext{1}{Number of antennas;}
\tablefoottext{2}{sideband containing the $^{12}$CO emission line (given are lower sideband (left) and upper sideband (right) rms).}}
\label{Obs-para}
\end{table*}

\object{HE 0433-1028} is a barred Seyfert 1 galaxy \citep{Kewley2001} at a redshift of $z = $0.03555 $\pm$ 0.00001 \citep{Keel1996}. On the one hand, it can be classified as a broad-line Seyfert 1 (BLS1) based on the Balmer line widths of more than 3000 km s$^{-1}$ \citep{Wang2007,Mullaney2008}, and, on the other hand, it can be assigned to the group of narrow-line Seyfert 1 (NLS1) with regard to its 
[OIII$] \lambda$5007/H$\beta$ line ratio \citep{Ryan2007,Mullaney2008}. The emission is dominated by the AGN component \citep{Yuan2010}. 
Up to now the host morphology and properties have not been a subject of discussion in the literature.

\object{HE 1029-1831} is a barred galaxy with two spiral arms and is located at a redshift of $z = $ 0.04026 $\pm$ 0.00009 \citep{Kaldare2003}. The AGN classification of HE 1029-1831 seems to be ambiguous: from the optical line ratio diagnostics, \citet{Kewley2001} determine that it ranges in the field between an \ion{H}{ii}/borderline and an AGN/borderline galaxy. However, \citet{Nagao2001} and \citet[][both optical]{RodArd2000}, and \citet[][NIR]{Fischer2006}
classify HE 1029-1831 as a NLS1 galaxy based on the Balmer, Pa$\alpha$, and Br$\gamma$ line widths of about 2000 km s$^{-1}$. The NIR line and color diagnostics indicate a mixture of starburst and AGN excitation \citep{Fischer2006,Yuan2010,Busch2014}. 

In \citet{Busch2014}, we find this galaxy
to have circumnuclear ring with two intense, but decreasing starburst regions containing an intermediate-age stellar population of $\sim$ 100 Myr in age. This population is likely to lower the mass-to-light ratio so that the galaxy follows the $M_\mathrm{BH}$--$M_\mathrm{bulge}$ relations but not the $M_\mathrm{BH}$--$L_\mathrm{bulge}$ relations of inactive galaxies \citep{Busch2014}.
Active star formation is also indicated by a FIR luminosity of $L_\mathrm{FIR}=1.9 \times 10^{11}~L_{\sun}$ \citep{Fischer2006} classifying it as a LIRG. 

The \ion{H}{i} gas mass is $M_\mathrm{\ion{H}{i}}=6.6 \times 10^{9} M_{\sun}$ \citep[][]{Konig2009}. 
The $^{12}$CO emission extends along the optical bar and shows a strong velocity gradient perpendicular to this bar,
which suggests a bar-driven inflow \citep[IRAM, Plateau de Bure Interferometer (PdBI) and the Berkeley-Illinois-Maryland Association (BIMA) observatory data,][]{Krips2007}. The $^{12}$CO (2--1)/(1--0) line ratio indicates that most of the molecular gas is cold and subthermally excited except from the southern part of the bar emission, which 
might correspond to the crossing point of the bar end with the spiral arm \citep[][]{Krips2007}.
The potential existence of warmer and denser gas in this region needs to be studied further. 

\object{HE 1108-2813} is a BLS1 galaxy with a redshift of $z = $ 0.02401 $\pm$ 0.00004 \citep{Theureau2005} and with Balmer broad line widths of about 5000 km s$^{-1}$ \citep{Kewley2001,Crenshaw2003,Wang2007}.
\citet{Kollatschny1983} classify it as Seyfert 1.5 galaxy based on the Balmer line peak ratio of the broad and narrow line components. They find the nucleus to have a steep (negative) continuum slope toward the UV, corresponding to the absence of higher ionization lines (e.g., \ion{C}{iv}, \ion{C}{ii}]), on the one hand, but strong optical and UV \ion{Fe}{ii}]) multiplets on the other. 

The faint host \citep[see, e.g., dominant AGN component;][]{Yuan2010} is a barred and dusty grand design spiral galaxy \citep{Deo2006} of Hubble type SBc/d \citep{Malkan1998} and seems to be bulgeless \citep{Orban2011}. The dust is mainly distributed in chaotic regions in the central kiloparsec and the bar, and in large scale lanes, along the leading edges of the bar and in the main spiral arms. Prominent star-forming regions are found along the bar edges \citep{Deo2006}. 


The paper is structured as follows. The observational setup, calibration procedure, and mapping characteristics are described in Section \ref{sec:obs}. Section \ref{sec:Results} covers the results from the observation, i.e., the CO emission spectra (\ref{subsec:spectra}), CO fluxes, gas masses and sizes (\ref{subsec:mass-size}), a detailed discussion of the morphology and the kinematics (\ref{subsec:morph-kin}) in all three galaxies, the dynamical mass estimate (\ref{sec:Mdyn}), and the dust properties (\ref{sec:dust-prop}). In Section \ref{sec:SF}, we assess the star formation properties. This is followed by a discussion of the properties of the interstellar matter based on the $^{12}$CO/$^{13}$CO(2--1) and/or $^{12}$CO $J+1 \leq 3$ line ratios and the morphology in Section \ref{sec:ISM}. Section \ref{sec:sum} summarizes the results of our study.


\begin{table}[!bp]
\centering
\caption{Calibrators}
\begin{tabular*}{0.485\textwidth}{@{\extracolsep{\fill}} cccc}
\toprule
Object &  Bandpass & Gain & Flux \\ 
\midrule
HE 0433-1028 & 3c454.3          & 0457-234      & Uranus        \\
                & (3c84)        &               &               \\
\\
HE 1029-1831 & 3c273    & 1058+015      & (Titan)       \\
                &               & 1130-148      & 1130-148      \\
\\
HE 1108-2813 & 3c273    & 1037-295      & Titan         \\ 
                & (3c84)        & 1058+015      & Ganymede      \\
\bottomrule
\end{tabular*}
\label{Calis}
\end{table}

\section{Observation and data reduction}
\label{sec:obs}

The three sources were observed with the SMA in Hawaii in compact configuration between 2007 and 2008. The two 2 GHz sidebands were placed such that they cover the $^{12}$CO(2--1) or (3--2) line transition, respectively, (see estimated rest frequencies in Table \ref{Obs-para}) in the one sideband and the weaker emission lines of $^{13}$CO(2-1) 
or HCO$^+$(4-3), respectively, in the other sideband to probe their fluxes. 
Each sideband contains 24 partially overlapping spectral windows with a bandwidth of 10 MHz. The spectral windows are subdivided into 128 channels with a resolution of 0.8125 MHz corresponding to 1.10 km s$^{-1}$ at 222.6 GHz and 0.73 km s$^{-1}$ at 332.4 GHz. HE 0433-1028 and HE 1029-1831 were observed for $\sim$ 2.5 hours in total 
and HE 1108-2813 was observed for $\sim$ 5 hours in total 
(see Table \ref{Obs-para}). The dish size of 6 m yields a field of view (FOV) of 56$''$ at 222.6 GHz and 37$''$ at 332.4 GHz. The spatial resolutions range from 1.5$''$ to 4.5$''$ (Table \ref{Obs-para}).


Table \ref{Calis} lists the observed calibrators.
Each on-source scan took 20 - 30 min and is framed by $\sim$ 4 min observations of the gain calibrators.
Except for the calibration of the system temperatures with the SMA version of the multichannel image reconstruction, image analysis, and display software package \citep[SMA MIRIAD,][]{Zhao2013}, all data sets were reduced and mapped with the Common Astronomy Software Application \citep[CASA, Version 3.2/3.3;][]{McMullin2007}. 
We used planets or their satellites as flux calibrators, except for HE 1029-1831, where neither Titan nor 3c273 yielded reliable flux values so that we use the gain calibrator's flux. The calibrated fluxes of all other calibrators are consistent within 20\% with the values found in the Submillimeter Calibrator List of the SMA Observer Center \footnote[1]{http://sma1.sma.hawaii.edu/smaoc.html}.

%
Our naturally weighted data cubes, with 256 $\times$ 256 pixels and a pixel scale of 0.2 $''$ pixel$^{-1}$, cover a velocity range from -200 to 200 km s$^{-1}$ with a channel resolution of 5 km s$^{-1}$.
The noise in these channels is less than 33 mJy at $^{12}$CO(2--1) and 63 mJy at $^{12}$CO(3--2) (see Table \ref{CO-lum-masses}).
The noise in the continuum is less than 1.5 mJy at 222 GHz and 3.9 mJy at 333 GHz (see Table \ref{Obs-para}).
For the moment maps we clipped the HE 1108-2813 $^{12}$CO(2--1) data cube at 3$\sigma_{\textrm{rms-ch}}$, and $\sigma_{\textrm{rms-ch}}$ is the channel rms noise $\sigma_{\textrm{rms-ch}}$ in Jy beam$^{-1}$.
All other data cubes are clipped at 2$\sigma_{\textrm{rms-ch}}$ since they contain a significant flux contribution below a 3$\sigma_{\textrm{rms-ch}}$ level. The difference in the velocity-integrated flux integrated over the source size for a 2- and 3$\sigma_{\textrm{rms-ch}}$-clipped cube is 10\% for $^{12}$CO and up to 40\% for $^{13}$CO.
All maps are primary beam corrected. 
The upper flux limit of an undetected line (nondetection) is calculated as three times $\sigma_{\textrm{rms-int}}$ in Jy beam$^{-1}$ km s$^{-1}$, where
$\sigma_{\textrm{rms-int}}=\sigma_{\textrm{rms-ch}}\sqrt{v_{\textrm{FZWI}} \; v_{\textrm{ch}}}$,
with 
$v_{\textrm{FZWI}}$ and $v_{\textrm{ch}}$ as the full width of the line at zero intensity (FWZI) and the width of the channel in km s$^{-1}$ \citep{Ivison1996}.
The upper limit of the continuum emission is given by three times the continuum noise level $\sigma_{\textrm{rms-cont}}$. The error of the integrated flux within a given region depends on the zero level error and the noise error in the integrated area, 
$\Delta S_{\textrm{tot}}=\sigma_{\textrm{rms-int}} \, m \, \sqrt{1/n + 1/m}$
, with $m$ and $n$ as the integrated area and the rms probe area divided by the beam area \citep{Klein1981,Andernach1999}.

To obtain information on the gas distribution on smaller spatial scales, we weighted the visibilities for the $^{12}$CO emission uniformly and applied a circular restoring beam. The beam size is given by the geometric average of the uniformly weighted beam axes. In the case of HE 1108-2813, it is given by the minor axis size of the uniformly weighted beam since we can assume the CLEAN components mimic the true brightness distribution very well as a result of the good SNR.


\section{Results}       
\label{sec:Results}

$^{12}$CO emission is clearly detected in all sources, whereas $^{13}$CO is tentatively detected in HE 0433-1028 and HE 1108-2813. HCO$^+$ line emission in HE 1029-1831 has not been detected. For the continuum emission in all three galaxies and the HCO$^+$ line emission in HE 1029-1831, we give upper limits. We refer to the moments 0, 1, and 2 of an image cube as the total/integrated flux, velocity field, and velocity dispersion, respectively. 

\begin{figure}[!htbp]                                                     
\centering $                                                            
\begin{array}{c}             
\vspace{0.1cm} 
\includegraphics[trim = 13mm 65mm 22mm 75mm, clip, width=0.47\textwidth]{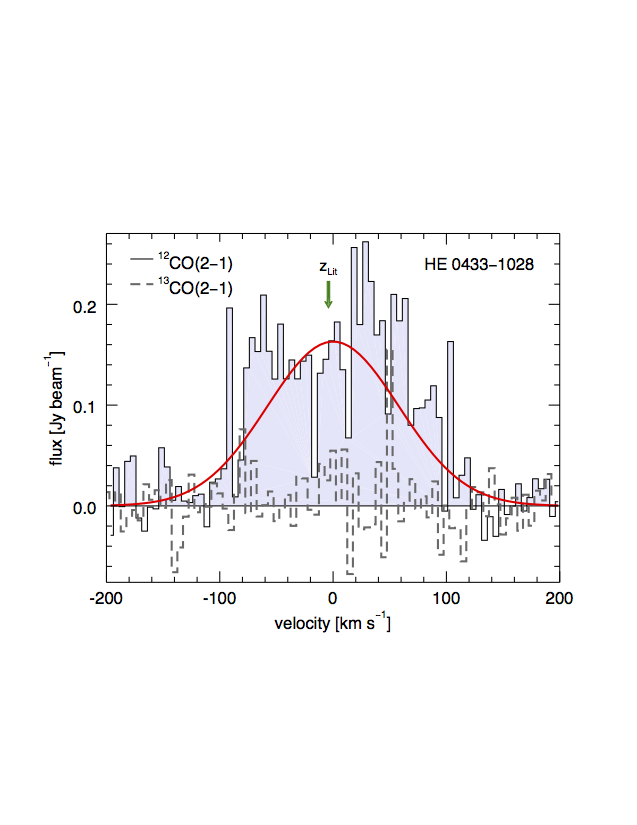}\\
\includegraphics[trim = 13mm 65mm 22mm 75mm, clip, width=0.47\textwidth]{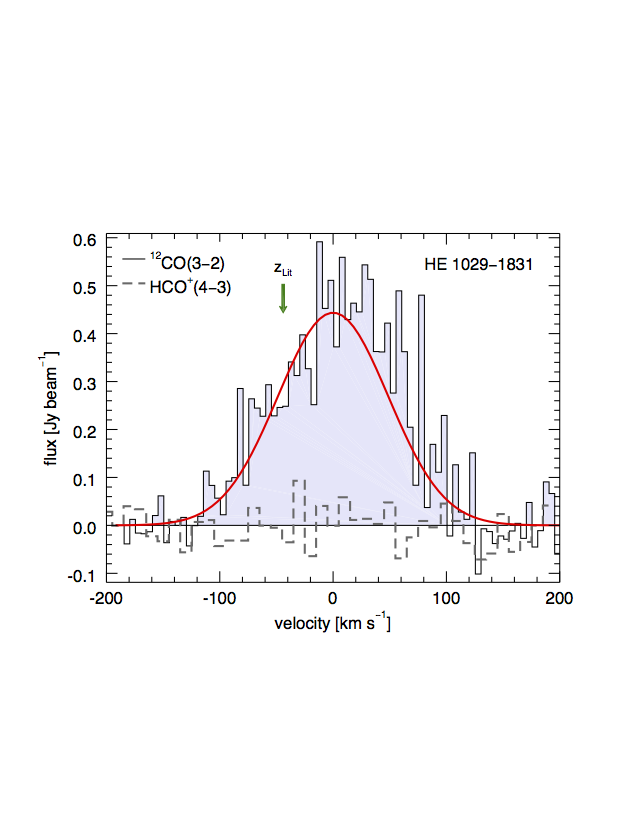}\\
\includegraphics[trim = 13mm 65mm 22mm 75mm, clip, width=0.47\textwidth]{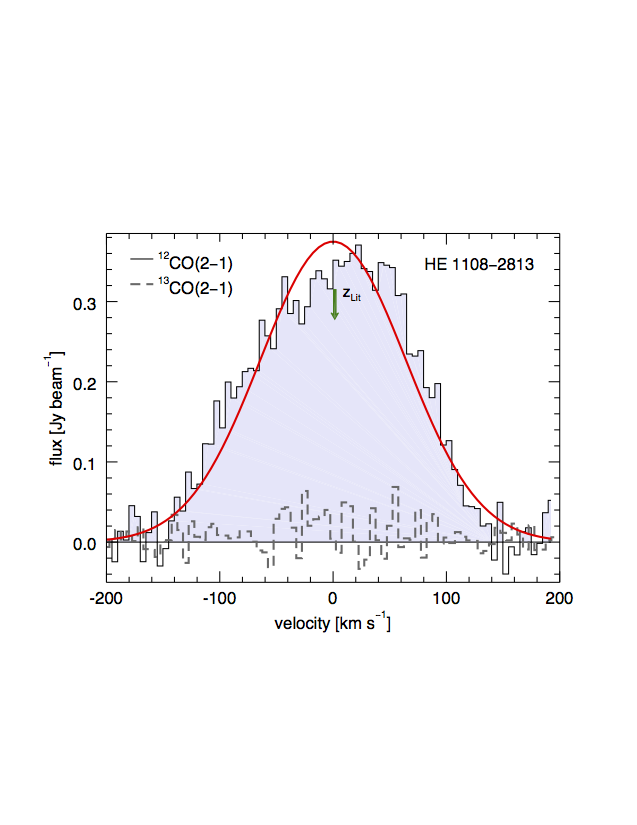}
\end{array} $
\caption{$^{12}$CO spectra (solid black line) of the three sources obtained from a region with the size of the beam centered at the position of the integrated flux peak (velocity in LSR). The (green) arrows indicate the optical redshift from the literature (see Table \ref{lit_props}). The parameters of the fitted Gaussian profile (solid red line) are given in Tables \ref{CO-spectra-Gauss-para} and \ref {CO-lum-masses}. In addition, the $^{13}$CO spectra of HE 0433-1028 and HE 1108-2813 from a beam-sized region at the position of the integrated $^{13}$CO flux peak and a HCO$^+$ spectrum (nondetection) of HE 1029-1831 at the position of the integrated $^{12}$CO flux peak are shown (gray dashed line).
}
\label{CO-spectra}
\end{figure}

\subsection{CO emission spectra}
\label{subsec:spectra}

We extracted spectra (Fig. \ref{CO-spectra}) from a region of the size of the beam and centered on the peak of the integrated $^{12}$CO emission. Although there are slight signs of skewness in the line profiles of all galaxies, and hints of a double horn profile in the case of HE 0433-1028, we chose to fit a Gaussian profile to obtain general characteristics of the $^{12}$CO emission listed in Table \ref{CO-spectra-Gauss-para}. Thus, the errors of the Gaussian parameters might be larger by a factor of 5 to 10, i.e., a redshift error of 0.0001. 
The found redshifts are consistent with literature (Table \ref{lit_props}) within their uncertainties. 
The full width at half maximum ($FWHM$) values range between 115 - 150 km s$^{-1}$. The $FWZI$ (Table \ref{CO-lum-masses}), given by the number of image cube channels in which emission is detected, are about 200 - 260 km s$^{-1}$. The $FWZI$ measured by \citet{Bertram2007} are about 50 - 60 km s$^{-1}$ larger for HE 1029-1831 and HE 1108-2813, but consistent for HE 0433-1028. The emission might be resolved out by the interferometer.
In addition, we extracted the spectra (Fig. \ref{CO-spectra}) around the $^{13}$CO line frequency of HE 0433-1028 and HE 1108-2813 and of the HCO$^+$ line frequency of HE 1029-1831. Again, we used a region with the size of the beam and centered it on the integrated $^{13}$CO emission peak in the case of the first two galaxies and on the integrated $^{12}$CO flux peak for the latter. While hints of emission in $^{13}$CO in HE 0433-1028 and HE 1108-2813 are slightly recognizable, i.e., well above the rms-noise of 1$\sigma_{\textrm{rms-ch}}$ (Table \ref{Obs-para}) especially for the latter, HCO$^+$ emission is clearly not detected in HE 1029-1831 despite the doubled channel size of 10 km s$^{-1}$ with a corresponding rms noise of 1$\sigma_{\textrm{rms-ch}}$ = 46 mJy beam$^{-1}$.


\begin{table*}[!htbp]
\centering
\caption{Position of the peak in integrated $^{12}$CO flux and parameters of the Gaussian profile fitted to the spectra at that position}
\tabcolsep=0.11cm
\begin{tabular*}{\textwidth}{@{\extracolsep{\fill}} ccccccccc}
\toprule
Object          & $^{12}$CO &\multicolumn{2}{c}{Peak position}\tablefootmark{1}         & $S_\textrm{peak}$     & Line center $\nu_\textrm{obs}$        & Line width $\sigma$     & $FWHM$                &       z       \\
                & transition & $\Delta \alpha$ [$''$]& $\Delta \delta$ [$''$]   & [mJy beam$^{-1}$]       &       [GHz]           &  [GHz]                & [km s$^{-1}$]   &               \\
\midrule                                                                                                                                                
HE 0433-1028    & 2--1  & 1.70                  & -1.00                         & 163 $\pm$ 13            & 222.621 $\pm$ 0.004   & 0.043 $\pm$ 0.003     & 137 $\pm$ 9     & 0.03556 $\pm$ 0.00002\\                                                                                                                            
HE 1029-1831    & 3--2  & 0.20                  &  0.20                         & 444 $\pm$ 28            & 332.363 $\pm$ 0.003   & 0.054 $\pm$ 0.003     & 115 $\pm$ 6     & 0.04041 $\pm$ 0.00001\\                                                                                                                            
HE 1108-2813    & 2--1  & 0.18                  & -0.41                         & 375 $\pm$ 17            & 225.133 $\pm$ 0.002   & 0.048 $\pm$ 0.001     & 151 $\pm$ 4     & 0.02401 $\pm$ 0.00001\\
\bottomrule
\end{tabular*}
\tablefoot{
\tablefoottext{1}{Offset relative phase tracking center.}}
\label{CO-spectra-Gauss-para}
\end{table*}

\begin{table*}[!htbp]
\centering
\caption{Source fluxes, line ratios, luminosities, and gas masses of the sources using the redshifts listed in Table \ref{CO-spectra-Gauss-para}.}\tabcolsep=0.10cm
\begin{tabular*}{\textwidth}{@{\extracolsep{\fill}} cccccccccc}
\toprule
Object          & Line                  & $FWZI_\textrm{im}$    & $S_\textrm{rms}$ &               $S_\textrm{peak~value}$         & $S_\textrm{tot}$\,($r\leq10''$) \tablefootmark{1}       & $L'_\textrm{$^{12}$CO(X--Y)}$         & \multirow{2}{*}{$\frac{\textrm{$^{12}$CO(X--Y)}}{\textrm{$^{12}$CO(1--0)}}$}  & \multirow{2}{*}{$\frac{\textrm{$^{12}$CO(X--Y)}}{\textrm{$^{13}$CO or HCO$^+$}} $}      & $M_{\textrm{mol}}$\tablefootmark{4}    \\                     
                &                       & [km\,s$^{-1}$] & \multicolumn{2}{c}{[Jy\,beam$^{-1}$\,km\,s$^{-1}$]}  & [Jy\,km\,s$^{-1}$]                                              &  [$10^8$\,K\,km\,s$^{-1}$\,pc$^2$]    &                                                 &   & [$10^9 M_\odot$]  \\                      
\midrule  
HE\,0433-1028   & $^{12}$CO(2--1)       &       195     & 0.9 &       33.2 &          $\;\,$56.6 $\pm$ 3.3 & 8.19 & 0.63 \tablefootmark{2}  &               & 1.0 - 6.2 \\                  
                & $^{13}$CO(2--1)       &       150     & 0.8 &  $\;\,$3.8 &       $\;\,\;\,$3.2 $\pm$ 1.2 &      &   -                     &  5 - 20       &           \\                  
HE\,1029-1831   & $^{12}$CO(3--2)       &       205     & 1.9 &       81.4 &               110.3 $\pm$ 6.1 & 9.17 & 1.00 \tablefootmark{3}  &               & 0.7 - 4.5 \\                  
                & HCO$^+$(4--3)         &    $\sim$ 150 & 1.8 &  $\;\,$ -  &         $\leq$ 5.3           &      &                         & $\geq$16          &           \\                 
HE\,1108-2813   & $^{12}$CO(2--1)       &       260     & 0.8 &       78.3 &               104.5 $\pm$ 3.1 & 6.84 & 0.65 \tablefootmark{2}  &               & 0.9 - 5.5 \\                  
                & $^{13}$CO(2--1)       &       150     & 0.5 &  $\;\,$3.5 &       $\;\,\;\,$3.0 $\pm$ 0.7 &      &                         &  5 - 30       &           \\                  
\bottomrule

\end{tabular*}
\tablefoot{
\tablefoottext{1}{Flux errors do not include the flux calibration error of 20\% - 30\% and are based on image noise to demonstrate the data quality only;}
\tablefoottext{2}{based on single dish fluxes in \citet{Bertram2007} and corrected for source size and our cosmology;}
\tablefoottext{3}{based on line ratio map of PdBI data \citep{Krips2007} and our data in Fig. \ref{HE1029-ratio};}
\tablefoottext{4}{total molecular mass in this observation (including Helium via a factor of 1.36), covering a mass range corresponding to $\alpha_\mathrm{CO} =$ 0.8 - 4.8 $M_\odot$~(K km s$^{-1}$ pc$^2$)$^{-1}$ for ULIRG and Galactic mass conversion \citep{Downes1998,Solomon1991}.}
}
\label{CO-lum-masses}
\end{table*}

\begin{table*}[!htbp]
\centering
\caption{Source distances, sizes (deconvolved Gaussian FWHM), and fractions $f_\textrm{Gauss}$ of the total flux contained in a Gaussian of given size.}
\begin{tabular*}{\textwidth}{@{\extracolsep{\fill}} ccccccc}    
\toprule
Object  & $D_\textrm{L}$\tablefootmark{1}  & Scale\tablefootmark{1} &   \multicolumn{2}{c}{$d_\textrm{FWHM$_1$} \times d_\textrm{FWHM$_2$}$}  & PA & $f_\textrm{Gauss}$ \\
        & [Mpc]           & [kpc/$''$]          & [$''~ \times~ ''$]     & [kpc $\times$ kpc] & [$^{\circ}$]     & [\%]  \\                                                                                                 
\midrule                                                                                                                        
HE 0433-1028 & 156.5 & 0.708 & 2.3 $\pm$ 0.2  $\times$  2.5$\pm$ 0.2 & 1.6 $\pm$ 0.1  $\times$  1.8 $\pm$ 0.1 & 31 & 94 \\ 
HE 1029-1831 & 178.4 & 0.799 & 1.3 $\pm$ 0.1  $\times$  1.3$\pm$ 0.2 & 1.1 $\pm$ 0.1  $\times$  1.1 $\pm$ 0.2 & - 25  & 96 \\ 
HE 1108-2813 & 104.7 & 0.484 & 3.3 $\pm$ 0.1  $\times$  1.4$\pm$ 0.1 & 1.6 $\pm$ 0.1  $\times$  0.7 $\pm$ 0.1 & 11 & 78 \\ 
\bottomrule

\end{tabular*}
\tablefoot{
\tablefoottext{1}{Obtained from Ned Wright's Cosmology Calculator \citet{NedWright2006} (http://www.astro.ucla.edu/$\sim$wright/CosmoCalc.html).}}
\label{CO-source-fit}
\end{table*}



\subsection{CO fluxes, gas masses, and sizes}
\label{subsec:mass-size}

\begin{figure*}[!htbp]
\centering $
\begin{array}{cc}
\includegraphics[trim = 1mm 0mm 6mm 2mm, clip, width=0.47\textwidth]{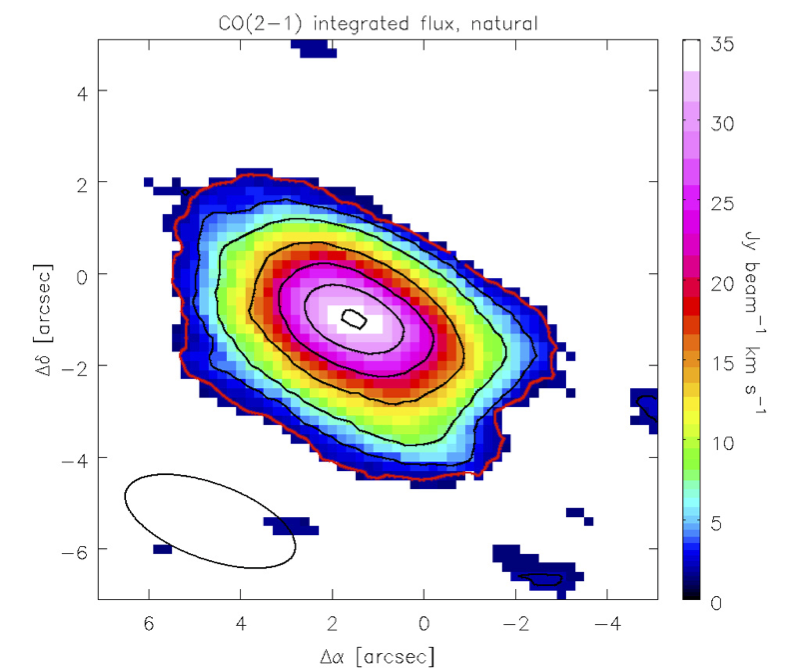}&
\includegraphics[trim = 1mm 0mm 6mm 2mm, clip, width=0.47\textwidth]{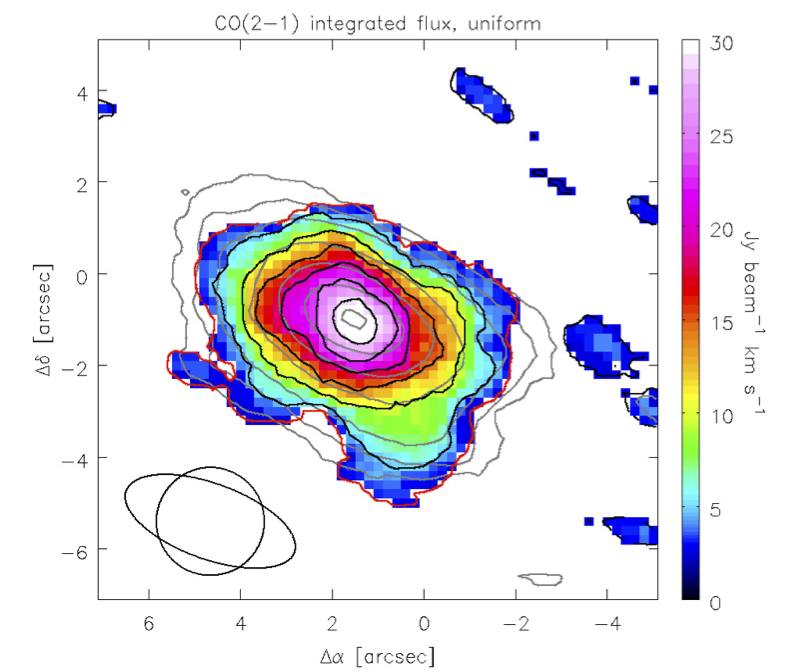}\\
\includegraphics[trim = 1mm 0mm 6mm 2mm, clip, width=0.47\textwidth]{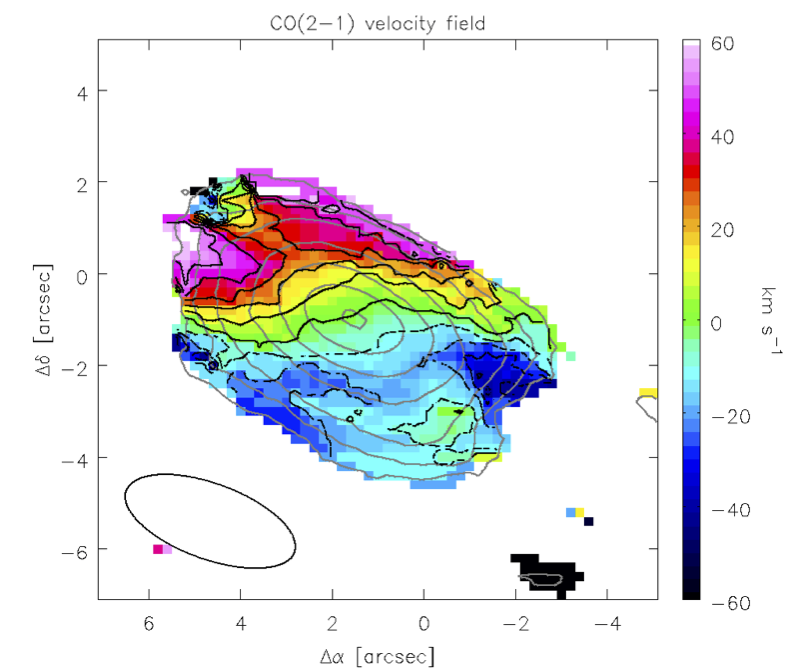}&
\includegraphics[trim = 1mm 0mm 6mm 2mm, clip, width=0.47\textwidth]{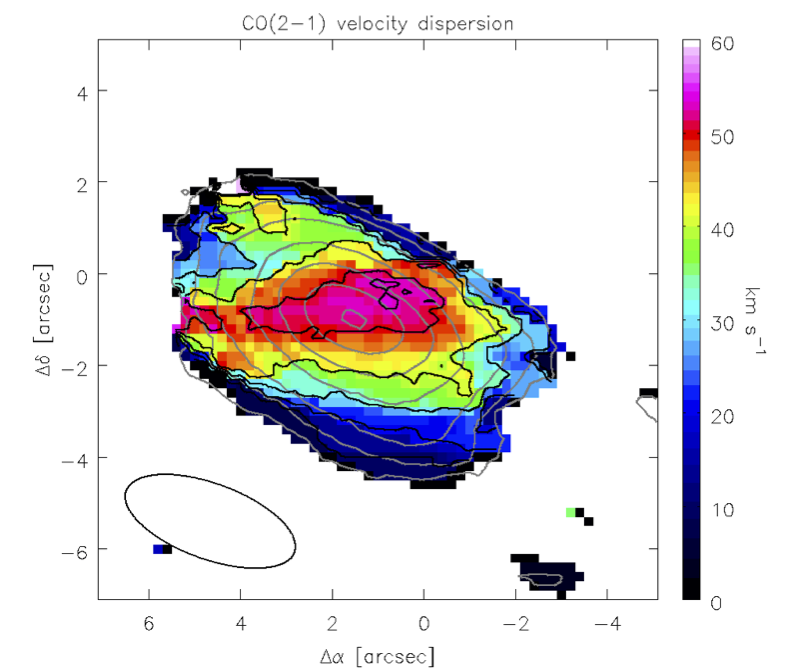}\\
\includegraphics[trim = 1mm 0mm 6mm 2mm, clip, width=0.47\textwidth]{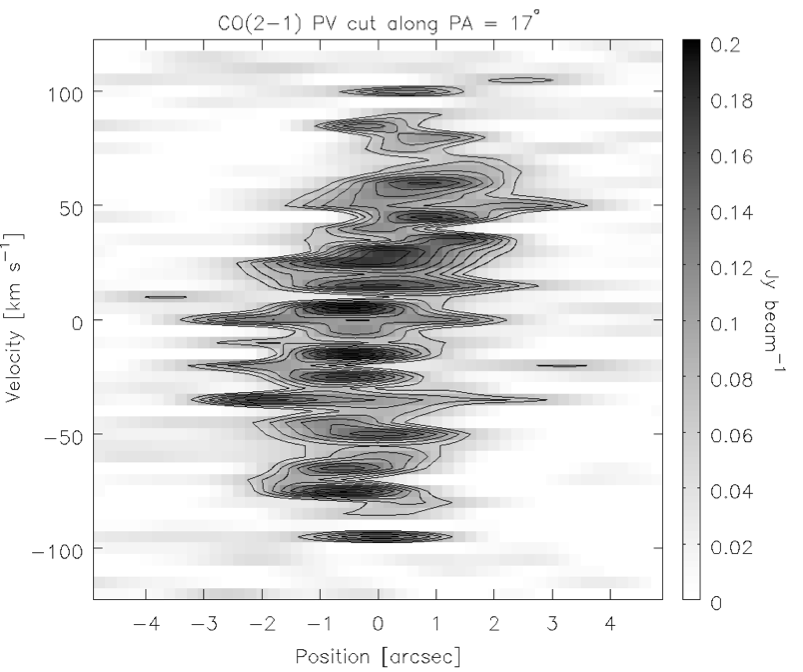}&
\includegraphics[trim = 1mm 0mm 6mm 2mm, clip, width=0.47\textwidth]{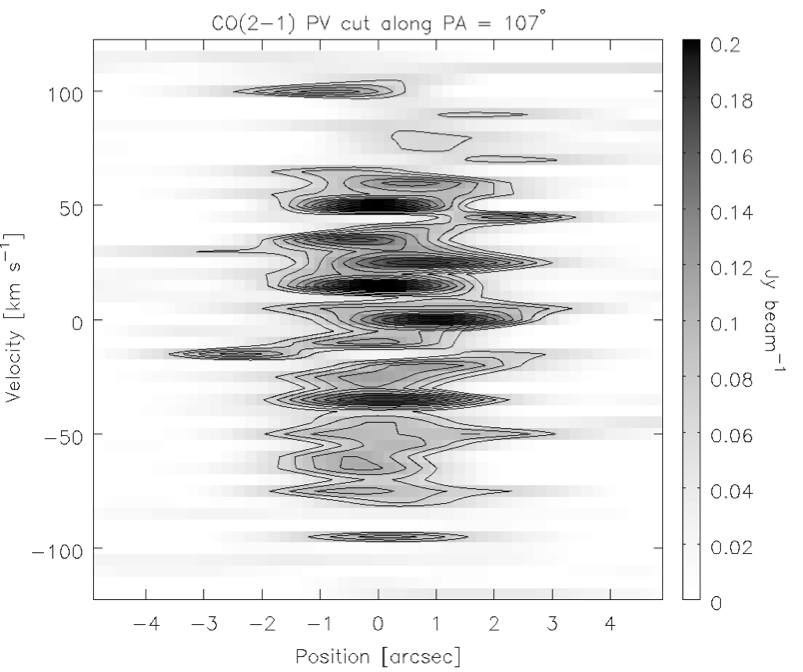}\\
\end{array} $                                                           
\caption{HE 0433-1028. 
\emph{Top panels:} the integrated $^{12}$CO(2--1) flux image for natural weighting (\emph{left}) and uniform weighting (\emph{right}) obtained for a velocity range from -100 to +100 km s$^{-1}$. The beam sizes are $3.9'' \times 1.6''$ and $2.35''$, respectively.
The contours are in steps of (2 (red), 4, 8, 16, 24, 32, 38) $\times$ 1$\sigma_{\textrm{rms-int}}$ (= 0.9 Jy beam$^{-1}$ km s$^{-1}$) and (2 (red), 4, 8, 12, 16, 20, 24) $\times$ 1$\sigma_{\textrm{rms-int}}$ (= 1.2 Jy beam$^{-1}$ km s$^{-1}$) , respectively.
\emph{Middle panels:} the corresponding isovelocity (\emph{left}) and velocity dispersion image (\emph{right}) for natural weighting. Black contours are in steps of 10 km s$^{-1}$ from -50 to +50 km s$^{-1}$ and from 10 to 50 km s$^{-1}$ (additional contour at 53 km s$^{-1}$), respectively. Gray contours show the naturally weighted integrated $^{12}$CO(2--1) flux.
\emph{Bottom panels:} the position velocity diagrams along the major (PA = 17$^\circ$, \emph{left}) and minor (PA = 107$^\circ$, \emph{right}) axis.
}
\label{HE0433-int-vel-disp}
\end{figure*}

The integrated flux maps, velocity fields, velocity dispersion maps, and position-velocity (PV) diagrams of the $^{12}$CO emission of the three galaxies are shown in the Figures \ref{HE0433-int-vel-disp}, \ref{HE1029-int-vel-disp}, and \ref{HE1108-int-vel-disp}.
The $^{13}$CO emission in HE 0433-1028 and HE 1108-2813 and the line luminosity ratios to the corresponding $^{12}$CO emission are shown in Fig. \ref{HE0433-ratio} and \ref{HE1108-ratio}, respectively.
Figure \ref{HE1029-ratio} gives the line luminosity ratios of the three lowest $^{12}$CO transitions in HE 1029-1831.
The source size of the dominating $^{12}$CO flux component are given in Table \ref{CO-source-fit}, and the integrated fluxes of all in this work detected lines are given in Table \ref{CO-lum-masses}.

The positions of the integrated flux peak deviate from the phase tracking centers less than 0.5$''$ (Table \ref{CO-spectra-Gauss-para}), except for HE 0433-1028, whose peak is located $\sim$ 2$''$ toward the southeast (Fig. \ref{HE1029-int-vel-disp}).
The overall size of the emission regions of HE 0433-1028, HE 1029-1831, and HE 1108-2813 extend across $\sim$ 9$''$, 5$''$, 18$''$ (= 6.4, 4.0, 8.7 kpc), respectively (see Figures \ref{HE0433-int-vel-disp}, \ref{HE1029-int-vel-disp}, and \ref{HE1108-int-vel-disp}). We fitted an elliptical Gaussian component to the dominating central emission region in the uv plane. 
They are listed in Table \ref{CO-source-fit}: about 80 - 95 \% of the total flux is concentrated within a deconvolved source size of $d_\textrm{FWHM} \lesssim$ 0.7 - 1.8 kpc. 
For HE 0433-1028 and HE 1108-2813, we cannot constrain properly how much flux has been resolved out compared to the single dish observations of \citet{Bertram2007}, since the primary bars are larger than the 11.3$''$ $^{12}$CO(2--1) beam of the IRAM 30m telescope and, for HE 0433-1028, even larger than the 22.7$''$ $^{12}$CO(1--0) beam. It is possible that not all flux is detected by the single dish measurement.
The bar sizes (major and minor axis) are estimated from optical images in Fig. \ref{HE0433-overlay} and \ref{HE1108-overlay} to be $29''\times 13''$ and $17.5'' \times 6''$, respectively. Furthermore, the emission is not homogeneous over the bar, but concentrated in the center. Hence, we construct a weighted source size based on the knowledge from the Gaussian component source fit in Table \ref{CO-source-fit}   
\begin{equation*}
\theta_\textrm{wgt-src}^2 = (1 - f_\textrm{Gauss}) \; \theta_\textrm{bar$_1$} \;\theta_\textrm{bar$_2$} + f_\textrm{Gauss} \; d_\textrm{FWHM$_1$} \; d_\textrm{FWHM$_2$}, 
\end{equation*}
with $\theta_\textrm{bar$_{1,2}$}$ and $d_\textrm{FWHM$_{1,2}$}$ denoting the major and minor axis of the bar and the fitted central component (Table \ref{CO-source-fit}).
We expect the $^{12}$CO(2--1) emission in HE 0433-1028 to trace cool gas and therefore to extend along the bar, similar to HE 1108-2813, but it is
undetected because of the three times lower SNR. Therefore, we assume the same flux fraction $f_\textrm{Gauss}$ for the central component like in HE 1108-2813, i.e., 78\% , in constrast to to our measurement. 
Applying the beam convolved source sizes $\theta_\textrm{conv}^2 = \theta_\textrm{wgt-src}^2 + \theta_\textrm{beam}^2 $ and our cosmology to the IRAM 30m fluxes \citep{Bertram2007}, we find for HE 0433-1028 that 61\% (37\% for $f_\textrm{Gauss}=1$) of IRAM 30m $^{12}$CO(2--1) flux is not detected by the SMA. As mentioned we cannot rule out a significant sensitivity issue against the spatial filtering of diffuse gas. 
In the case of HE 1108-2813, 88\% (103\% for $f_\textrm{Gauss}=1$) of the flux is recovered by the SMA.
In addition, the calibration errors (20 - 30\%) need to be kept in mind.
%
%
%
%
%
For HE 1029-1831, we have no literature values for comparison of the $^{12}$CO(3--2) emission at hand.

Following the formalism of \citet{Solomon1991}, we derive the $^{12}$CO(1--0) luminosities according to
\begin{equation*}
L'_{\textrm{CO}}=3.25 \times 10^7 \, S_{\textrm{CO}}\Delta V \; D_\textrm{L}^2  \; \nu_{\textrm{obs}}^{-2} \; (1+z)^{-3},
\end{equation*}
with the integrated flux $S_{\textrm{CO}}\Delta V \equiv S_\textrm{total}$ in Jy km s$^{-1}$, the luminosity distance in $D_\textrm{L}$ in Mpc, the observed frequency $\nu_{\textrm{obs}}$ in GHz, and $L'_\textrm{CO}$ in K km s$^{-1}$ pc$^{2}$.
%

To derive the gas masses that are defined by the $^{12}$CO(1--0) transition, we first determine the global line (luminosity) ratios (see Table \ref{CO-lum-masses}), i.e., 
$r_{21} = S_2/S_1 \cdot \nu_2^2/\nu_1^2 =  I_2/I_1 \; \cdot \; \theta_\textrm{conv$_2$}^2/\theta_\textrm{conv$_1$}^2$, 
with the indices 1 and 2 denoting two different line transitions, from the $^{12}$CO data from \citet{Bertram2007}, considering the weighted source sizes $\theta_\textrm{wgt-src}$ and our cosmology. 
The line ratio maps in Fig. \ref{HE0433-ratio}, \ref{HE1029-ratio}, and \ref{HE1108-ratio} are derived analogously to this formula.
%
The $^{12}$CO(2--1)/(1--0) line ratios of $r_{21} \leq$ 0.65 indicates almost subthermal emission in all sources. 
For bar axis error of 1$''$ and a deviation from $f_\textrm{Gauss}$ by 10 \%, the ratio errors are about 0.15. 
Assuming the single dish fluxes to be completely determined by either the central Gaussian component or, highly unlikely, the bar, we obtain a range of $r_{21} = 0.45 - 1.00$ and $r_{21} = 0.58 - 0.86$ for HE 0433-1028 and HE 1108-2813, respectively. 
The $^{12}$CO(3--2)/(1--0) line ratio of $r_{31} \sim$ 1 for HE 1029-1831 was obtained from a line ratio map with PdBI data (see Section \ref{subsubsec:morph-HE1029} and Fig. \ref{HE1029-ratio} for calculation and discussion) hinting at a thermalized $^{12}$CO(3--2) transition.

The molecular gas masses (including Helium) are calculated using the $^{12}$CO(X--Y)/(1--0) line ratio $r_\mathrm{X1}$ via 
\begin{equation*}
M_\textrm{CO(X--Y)} = \frac{\alpha_\mathrm{CO(1-0)}}{r_\mathrm{X1}} \; L'_{\textrm{CO(X--Y)}},
\end{equation*} 
and yield $ (0.7 - 6.2) \times 10^9 M_\odot$.
We cover the range of possible masses of the molecular gas using a conversion factor of $\alpha_\mathrm{CO} = 0.8 M_\odot$ (K km s$^{-1}$ pc$^2$)$^{-1} \equiv \alpha_\mathrm{ULIRG}$, typically used for ULIRGs \citep{Downes1998}, and $\alpha_\mathrm{CO} = 4.8 M_\odot$ (K km s$^{-1}$ pc$^2$)$^{-1} \equiv \alpha_\mathrm{MW}$, typical for the Milky Way \citep{Solomon1991}. These factors are discussed in Sect. \ref{sec:Xco}. 

The difference in the masses of HE 1029-1831 is striking:
Our $^{12}$CO(3--2) data trace 41 \% of the BIMA $^{12}$CO(1--0) detected mass, 58 \% of the PdBI $^{12}$CO(1--0) mass, but 98 \% of the PdBI $^{12}$CO(2--1) mass, i.e., $M_\textrm{CO(2--1)} =$ 4.5 $\times$ 10$^9 M_\odot$ ($\alpha_\mathrm{CO} = 4.8 M_\odot$ (K km s$^{-1}$ pc$^2$)$^{-1}$), using the $^{12}$CO(2--1) flux and the $^{12}$CO(2--1)/(1--0) line ratio of $\sim$ 0.5 in \citet{Krips2007}.
Missing flux and sensitivity might explain this discrepancy.
\citet{Monje2011} observed with the CSO 10.4 m telescope a $^{12}$CO(2--1) flux of $\sim$ 93 Jy km s$^{-1}$, suggesting 73 \% of the flux to be resolved out by the PdBI. The corresponding mass is 18 $\times$ 10$^9 M_\odot$ ($\alpha_\mathrm{CO} = 4.8 M_\odot$ (K km s$^{-1}$ pc$^2$)$^{-1}$), i.e., 1.6 times $M_\textrm{BIMA}$ for a $^{12}$CO(2--1)/(1--0) line ratio of $\sim$ 0.5. Assuming the CSO mass to be reliable the BIMA would have resolved out 40 \% of the flux. 
In addition, $^{12}$CO(3--2) traces the warmer and/or denser gas fraction, which is confined to the central region as suggested by the source extent compared to the whole galaxy (Fig. \ref{HE1029-overlay}). Hence, the deficit in traced mass can be explained by a mixture of excitation and spatial filtering effects.

Comparing the amount of molecular gas, local galaxies show masses around $10^8 -$ few $10^9 M_\odot$ \citep[e.g.,][]{Helfer2003,Israel2009}, whereas intermediate redshift (z $\sim$ 0.1 - 0.6) QSO hosts and ULIRGs have gas masses of few $10^8 -$ few $10^{10} M_\odot$ \citep{Combes2011,Krips2012,Villar-Martin2013,Rodriguez2014}. \citet{Tacconi2006} find the masses for their submillimeter galaxies sample (z $\sim$ 2) to range within $10^{10} - 10^{11} M_\odot$. The masses we obtained are rather large compared to local galaxies when using Galactic mass conversion and are rather small when comparing to intermediate redshift QSOs and using ULIRG mass conversion. 


\subsection{Morphology and kinematics}
\label{subsec:morph-kin}

In this section, we discuss the following features per galaxy: the morphology in naturally and uniformly weighted data, the kinematics from the velocity image, velocity dispersion image and PV diagrams, the distribution of $^{13}$CO and its luminosity line ratio with the $^{12}$CO, or in the case of HE 1029-1831 the luminosity line ratio with $^{12}$CO(2--1) and (1--0), and a comparison with optical or NIR data.
Beam smearing, especially in combination with the inclination, most likely affects the measured velocity dispersions because one line of sight includes gas moving at different velocities, leading to a broadening of the observed line profile beyond the gas intrinsic dispersion. Therefore, the dispersion and also maybe the velocity information need to be treated with caution.


\subsubsection{HE 0433-1028}
\label{subsubsec:morph-HE0433}

Figure \ref{HE0433-int-vel-disp} shows that the molecular gas is confined to a compact boxy region of irregular shape hinting at a complex morphology. The structure is partly resolved, but the major fraction of emission is unresolved at the center. 
In the uniformly weighted image, the complex morphology becomes more evident, and shows a central component and two east and west of it.

The velocity field covers a range from -50 to 55 km s$^{-1}$ and appears twisted toward the circumnuclear region. The central gradient is oriented at a PA = 48$^\circ$ (from north to east), whereas the outskirts can be described by a PA $\sim 0^\circ$. This twist in the isovelocity lines is indicative for a kinematically distinct feature, such as a secondary bar or an inclined nuclear disk. 
The dispersion peaks with $\sigma_\textrm{v}$ $\sim$ 55 km s$^{-1}$ at the center. The region of high dispersion extends along the 0 km s$^{-1}$ with a slight extension to the southwest, which coincides with a larger region of constant velocity (plateau).          
The edge of the emission region is likely to be noise contaminated.

The spatial distribution of velocities along an intermediate velocity gradient of PA $\sim$ 17$^\circ$ reveals an ambiguous structure: On the one hand, it can be approached by one rigid-rotation component (assumed for our dynamical mass estimate). On the other hand, it could represent two components (x-shape), as indicated by the velocity field, i.e., a steeper gradient for the central component, centered on 0$''$ offset and ranging from -100 to 100 km s$^{-1}$, and a smaller gradient, from -2.5$''$ to 2.5$''$ and -40 to 50 km s$^{-1}$, for the east and west components indicated in the uniformly weighted image. 
The PV cuts along the gradient of the center and outskirts did not provide a deeper insight.
The minor axis PV cut (PA $\sim$ 107$^\circ$) shows the same velocity range as the central velocity component in the x-shape case. Outliers at lower velocities stem from the edge of the emission region and are not reliable. 


\begin{figure}[!htbp]
\centering $
\begin{array}{c}
\includegraphics[trim = 1mm 0mm 6mm 2mm, clip, width=0.47\textwidth]{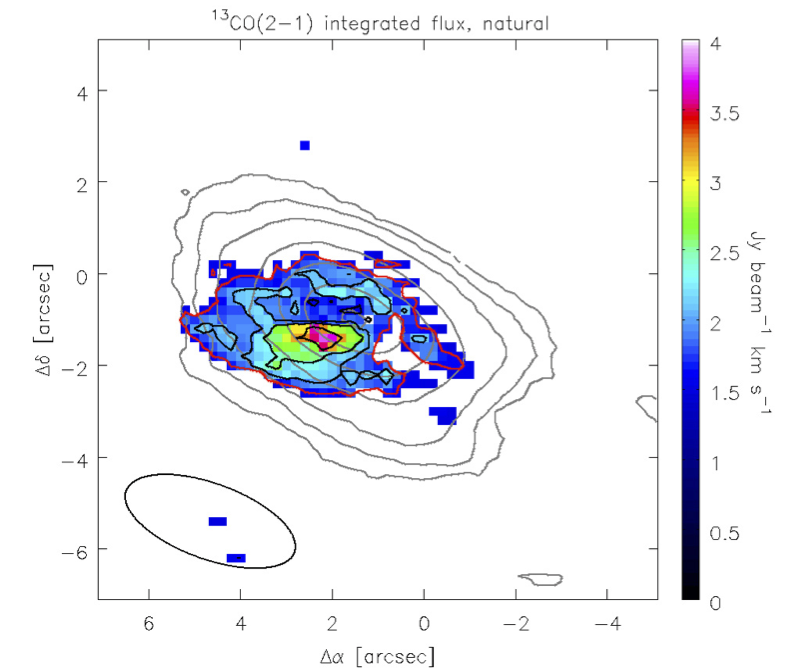}\\
\includegraphics[trim = 1mm 0mm 6mm 2mm, clip, width=0.47\textwidth]{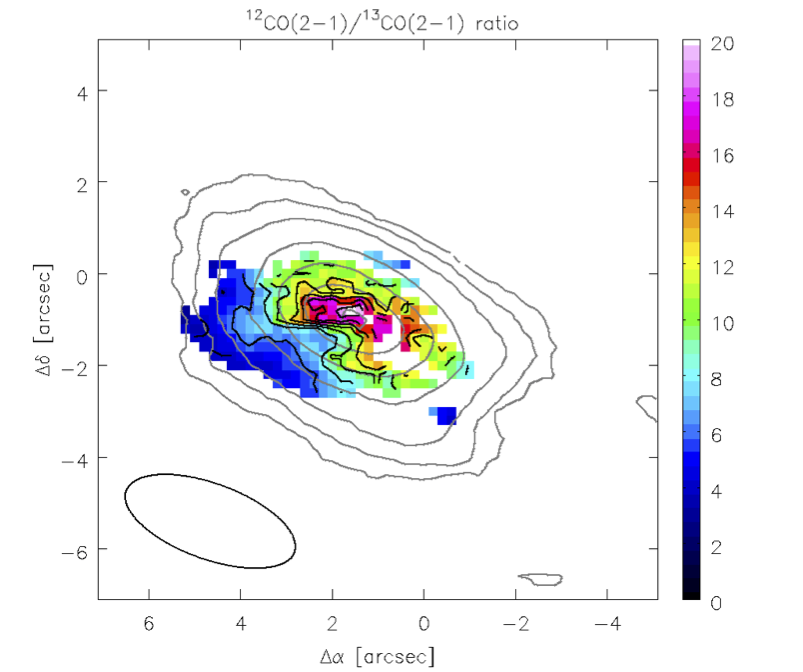}\\
\end{array} $                                                           
\caption{HE 0433-1028. 
\emph{Top:} the integrated $^{13}$CO(2--1) flux image for natural weighting obtained for a velocity range from -75 to +75 km s$^{-1}$. Black contours are in steps of (2 (red), 2.5, 3, 4) $\times$ 1$\sigma_{\textrm{rms-int}}$ (= 0.8 Jy beam$^{-1}$ km s$^{-1}$). The beam size is $3.9'' \times 1.6''$. 
\emph{Bottom:} the $^{12}$CO(2--1)/$^{13}$CO(2--1) luminosity ratio with black contours in steps of 2 from 4 to 18.
Gray contours show the naturally weighted integrated $^{12}$CO(2--1) flux as in Fig. \ref{HE0433-int-vel-disp}.
}
\label{HE0433-ratio}
\end{figure}

\begin{figure}[!htbp]
\centering $
\begin{array}{c}
\vspace{0.4cm} 
\includegraphics[trim = 0mm 1mm 50mm 1mm, clip, width=0.39\textwidth]{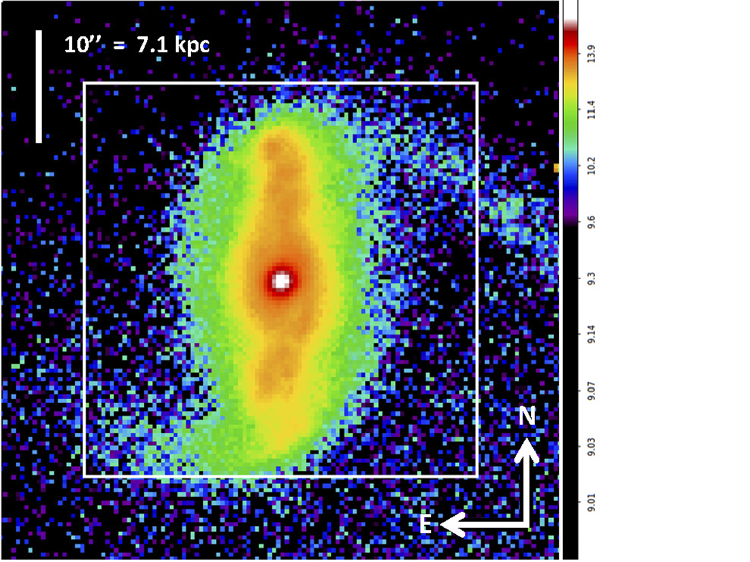}\\
\vspace{0.4cm}   
\includegraphics[trim = 0mm 1mm 50mm 1mm, clip, width=0.39\textwidth]{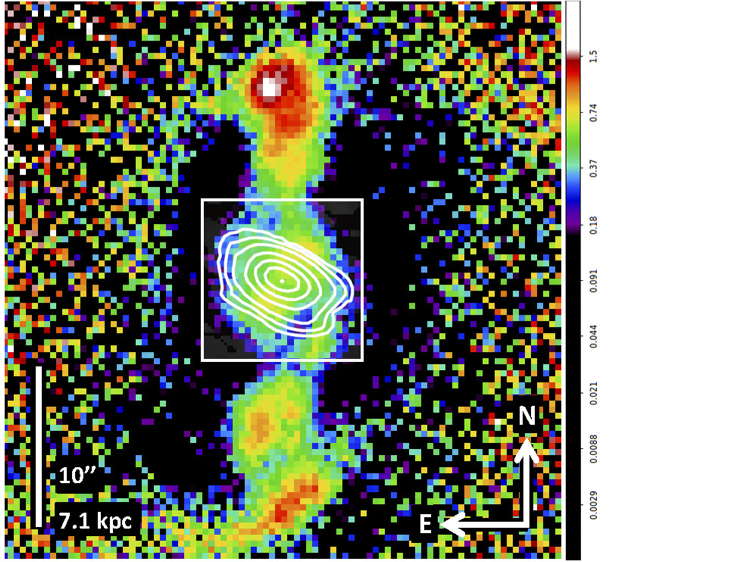}\\
\vspace{0.4cm}  
\includegraphics[trim = 0mm 1mm 50mm 1mm, clip, width=0.39\textwidth]{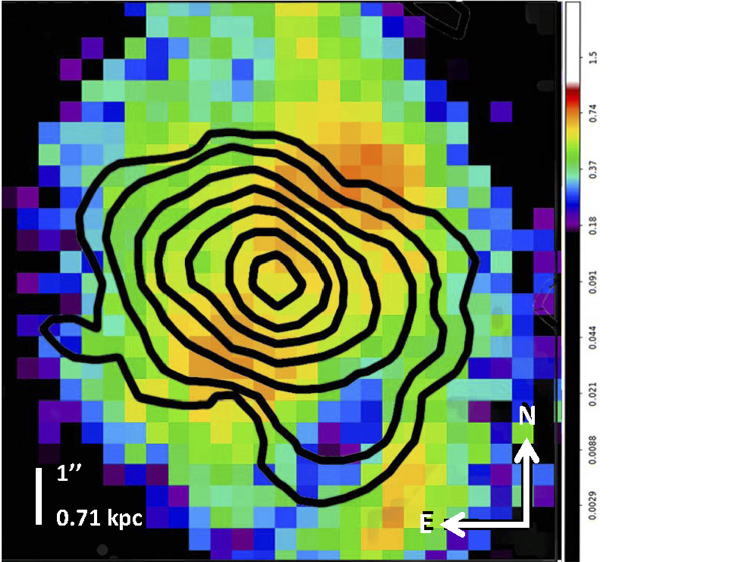}\\
\end{array} $                                                           
\caption{HE 0433-1028. 
\emph{Top:} $50'' \times 50''$ V-band image (ESO-Danish 1.54 m telescope, NED, arbitrary intensity units) of the galaxy, the box size is $35'' \times 35''$.
\emph{Middle:} 
zoomed into the box region of unsharp masked version of the image (arbitrary intensity units) overlayed with contours of naturally weighted integrated $^{12}$CO(2--1) flux as in Fig. \ref{HE0433-int-vel-disp}. Central region marked by box of size $10'' \times 10''$.
\emph{Bottom:} zoomed into inner $10'' \times 10''$ of unsharp masked image with contours of uniformly weighted integrated $^{12}$CO(2--1) flux as in Fig. \ref{HE0433-int-vel-disp}.
}
\label{HE0433-overlay}
\end{figure}


The $^{13}$CO emission (Fig. \ref{HE0433-ratio}) is contingently detected in the nuclear region. Its peak is slightly off-center ($\sim$ 0.5$''$, i.e., $<$ 0.5 $\times ~\theta_\textrm{minor}$) to the southeast, leaving a slight dip in emission in the center and there is more emission toward the east than the west. The $^{12}$CO/$^{13}$CO line ratio is 20 on center, 10 in a radius of r $\leq$ 2$''$ and 6 to the outskirts of the $^{12}$CO emission in the east.
We discuss the interpretation of these ratios later in Sect. \ref{sec:1213CO}.

To compare our data with the optical, we used a V-band (550nm) image (Fig. \ref{HE0433-overlay}) obtained with the ESO-Danish 1.54 m telescope \citep{Hunt1999}. This image
suggests the bar is clumpy and has a strong AGN component. We subtracted a Gaussian component fit with the dominating central component from the original and then unsharp-masked the Gauss-subtracted image to enhance the contrast of the clumps in the bar and at the center. Two bright peaks, on opposing sides of the circumnuclear region, as well as bright bar tips and dust lanes that wind toward
the nucleus, become visible. 
These features might indicate star formation sites in a circumnuclear ring or a secondary bar at PA = -37$^\circ$ $\pm$ 5$^\circ$. The latter matches our findings in the velocity distribution when we compare with \citet[][see their Fig. 11, middle panels]{Athanassoula2002}. 
A nuclear bar, misaligned with the global kinematic axes, seems to be slightly ahead of the twist in isovelocity contours in the center, in our case, $\sim 5^\circ$.

In an overlay with our uniformly weighted $^{12}$CO data, the molecular gas extends from the center out to the dust lane in the primary bar. The dust lanes appear to be pointing to the two clumps east and west of the central component in the uniformly weighted $^{12}$CO map (see also red scaled region in Fig. \ref{HE0433-int-vel-disp}, upper right panel), which are oriented at a PA = 74$^\circ$ $\pm$ 5$^\circ$.
These positions are also the location of plateaus in the naturally weighted (imaging) isovelocity map and local maxima in the dispersion (Fig. \ref{HE0433-int-vel-disp}, middle), indicating kinematics that are different from the central region. The extension in $^{12}$CO from the center toward the southwest is likely to be part of the southern dust lane.

\subsubsection{HE 1029-1831}
\label{subsubsec:morph-HE1029}

\begin{figure*}[!htbp]
\centering $
\begin{array}{cc}
\includegraphics[trim = 1mm 0mm 6mm 2mm, clip, width=0.47\textwidth]{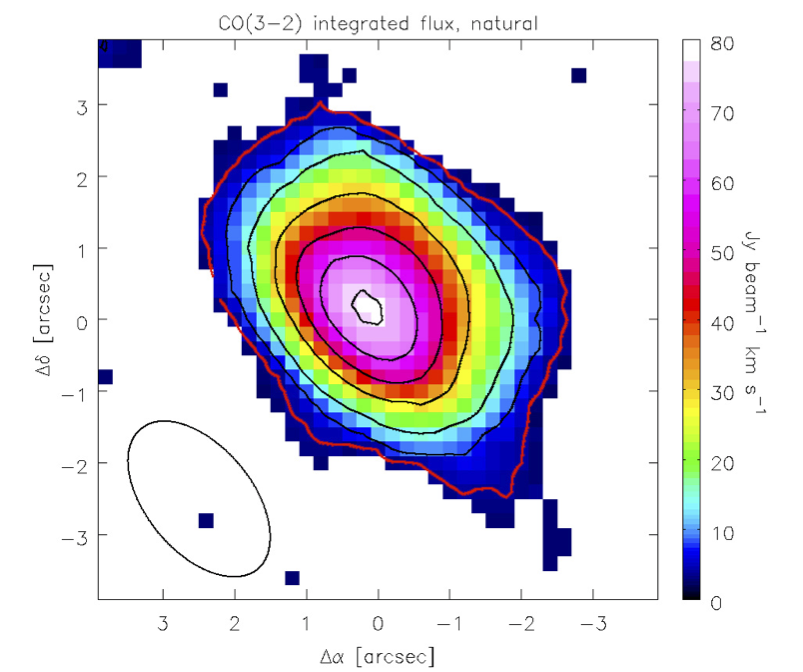}&
\includegraphics[trim = 1mm 0mm 6mm 2mm, clip, width=0.47\textwidth]{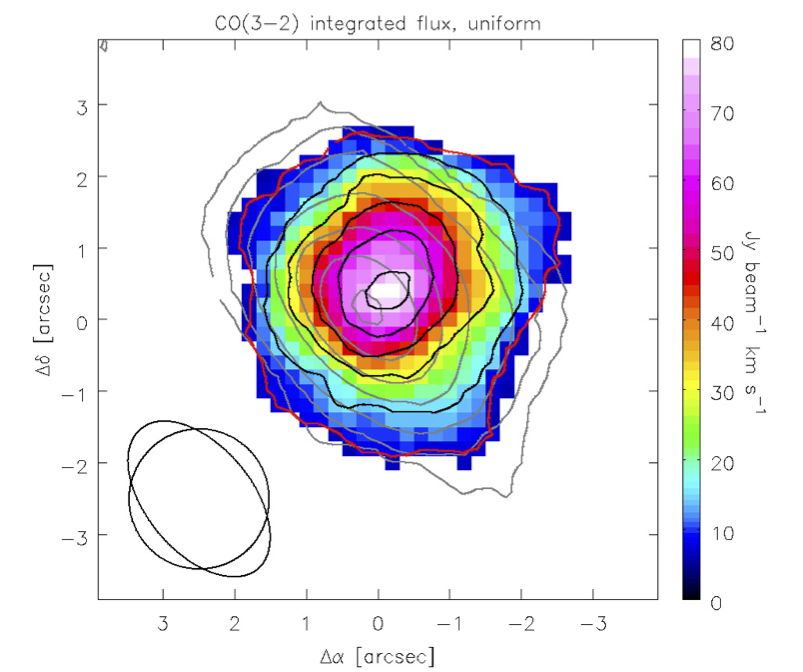}\\
\includegraphics[trim = 1mm 0mm 6mm 2mm, clip, width=0.47\textwidth]{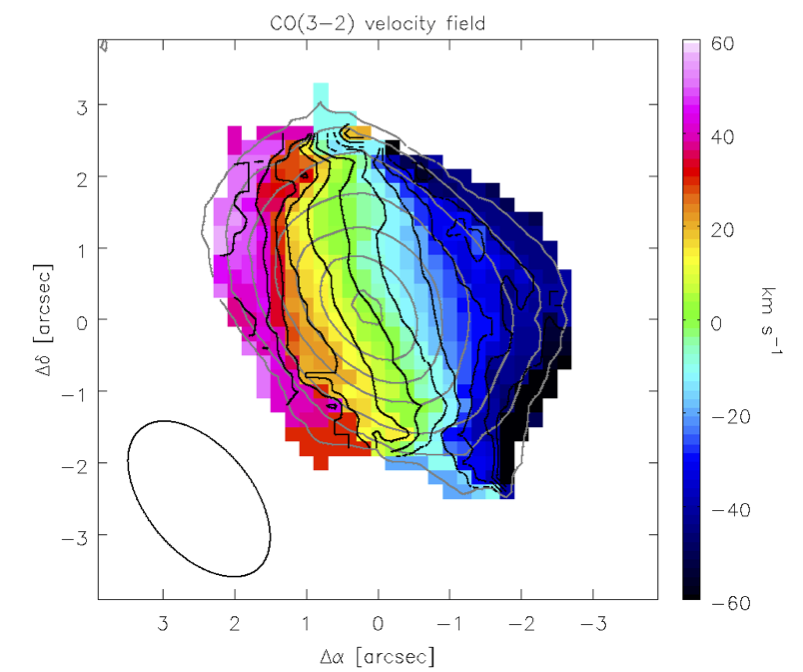}&
\includegraphics[trim = 1mm 0mm 6mm 2mm, clip, width=0.47\textwidth]{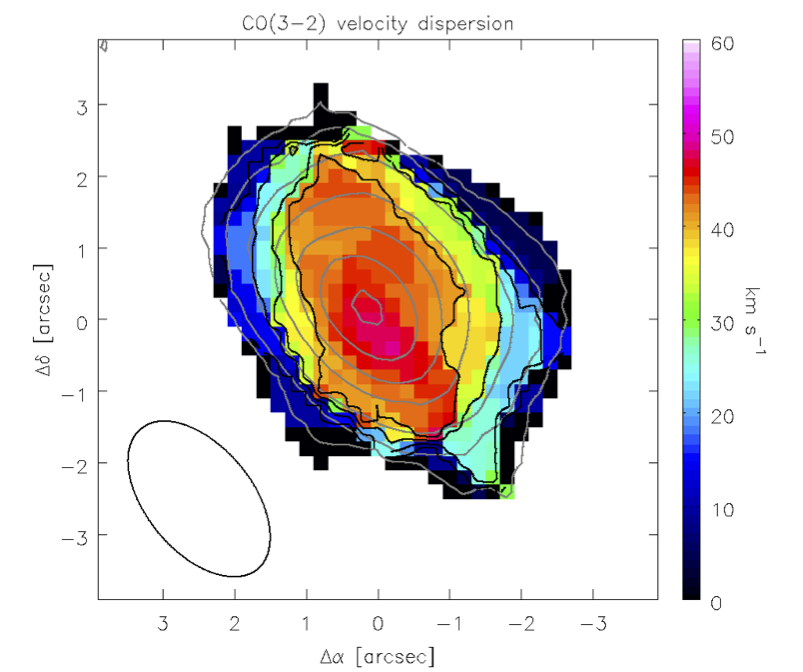}\\
\includegraphics[trim = 1mm 0mm 6mm 2mm, clip, width=0.47\textwidth]{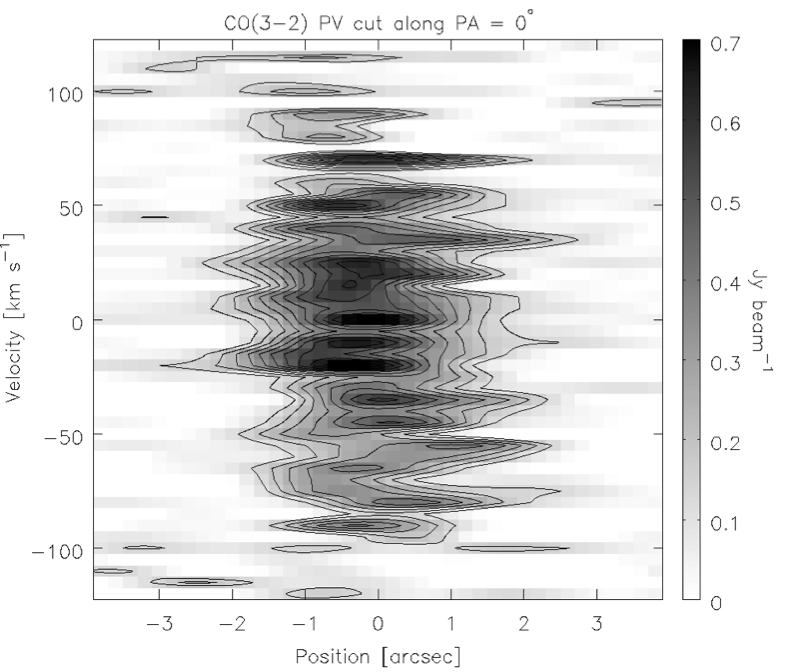}&
\includegraphics[trim = 1mm 0mm 6mm 2mm, clip, width=0.47\textwidth]{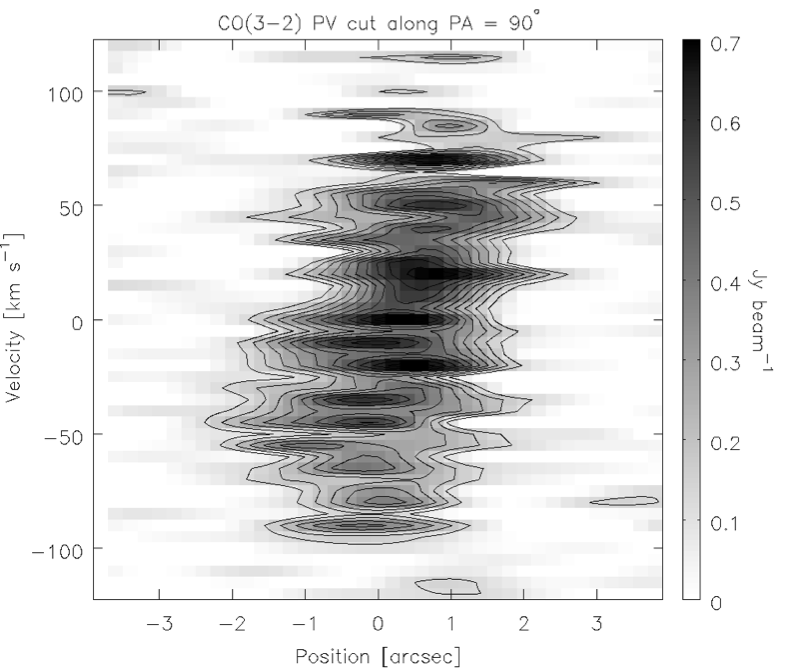}\\                                                                       
\end{array} $                                                           
\caption{HE 1029-1831.
\emph{Top panels:} the integrated $^{12}$CO(3--2) flux image for natural weighting (\emph{left}) and uniform weighting (\emph{right}) obtained for a velocity range from -100 to +100 km s$^{-1}$. The beam sizes are $2.5'' \times 1.5''$ and $1.95''$, respectively.
The contours are in steps of (2 (red), 4, 8, 16, 24, 32, 40) $\times$ 1$\sigma_{\textrm{rms-int}}$ (= 1.9 Jy beam$^{-1}$ km s$^{-1}$) and (2 (red), 4, 8, 12, 16, 20) $\times$ 1$\sigma_{\textrm{rms-int}}$ (= 3.7 Jy beam$^{-1}$ km s$^{-1}$), respectively.
\emph{Middle panels:} the corresponding isovelocity (\emph{left}) and velocity dispersion image (\emph{right}) for natural weighting. Black contours are in steps of 10 km s$^{-1}$ from -50 to +50 km s$^{-1}$ and from 10 to 50 km s$^{-1}$, respectively. Gray contours show the naturally weighted integrated $^{12}$CO(3--2) flux. 
\emph{Bottom panels:} the position velocity diagrams along the minor (PA = 0$^\circ$, \emph{left}) and major (PA = 90$^\circ$, \emph{right}) axis.
}   
\label{HE1029-int-vel-disp}
\end{figure*}

The $^{12}$CO(3--2) line emission (Fig. \ref{HE1029-int-vel-disp}) traces warm and dense molecular gas in a compact region, i.e., within a radius five times smaller than the $^{12}$CO(1--0) emission region observed by \citet{Krips2007}. The southern component detected by Krips et al. could not be verified, instead, we see a slight extension to the northwest (PA = -41$^\circ$ $\pm$ 5$^\circ$).
\citet{Krips2007} find the $^{12}$CO(1--0) and $^{12}$CO(2--1) line emission to extend even along the primary bar and the $^{12}$CO(2--1)/(1--0) line ratio of $\sim$ 0.5 suggests that most of the molecular gas is cold and subthermally excited.
In uniform weighting the slight asymmetry in the central region toward northwest (PA = -44$^\circ$ $\pm$ 5$^\circ$) becomes more evident. In fact, the peak position is shifted by 0.4$''$ (= 320 pc) toward the northwest, hinting at the presence of a second component apart from the central component.


\begin{figure*}[!htbp]
\centering $
\begin{array}{cc}
\includegraphics[trim = 1mm 0mm 6mm 2mm, clip, width=0.47\textwidth]{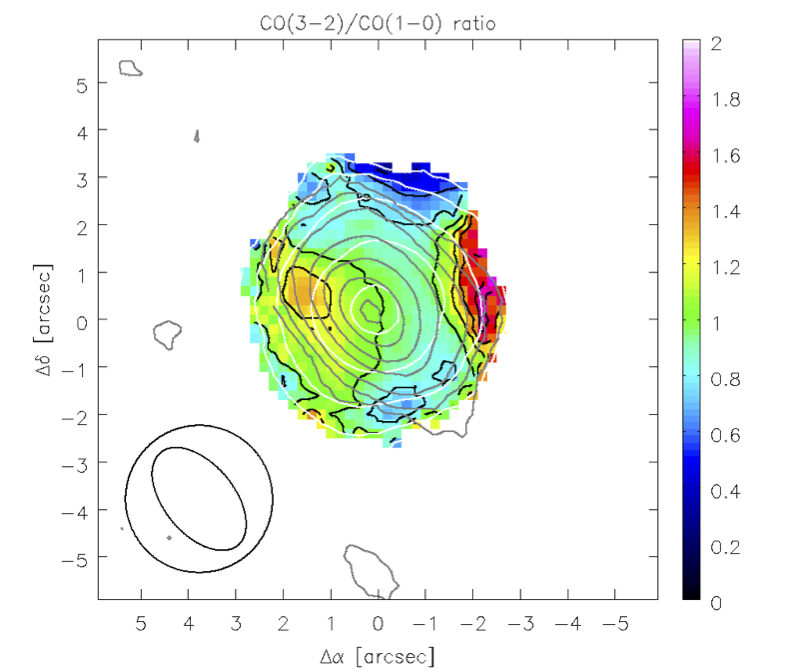}&
\includegraphics[trim = 1mm 0mm 6mm 2mm, clip, width=0.47\textwidth]{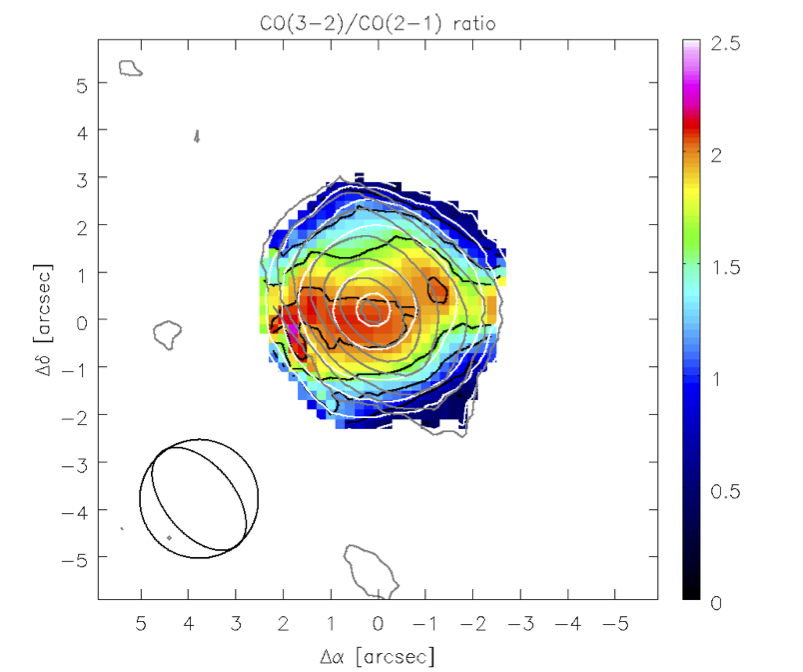}\\
\includegraphics[trim = 1mm 0mm 6mm 2mm, clip, width=0.47\textwidth]{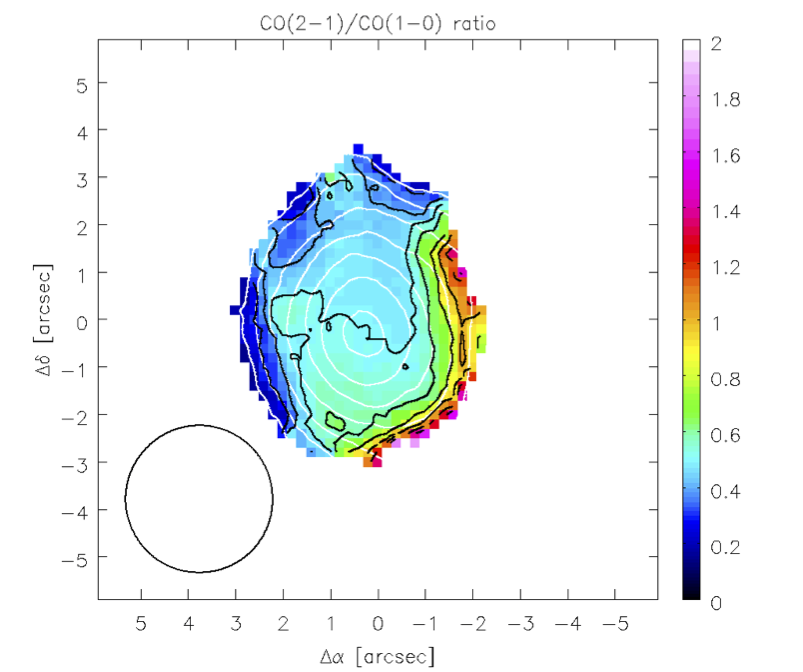}&
\begin{minipage}[b]{0.35\textwidth}
  \caption{HE 1029-1831.
\emph{Top left:} the $^{12}$CO(3--2)/$^{12}$CO(1--0) luminosity ratio with black contours in steps of (4, 6, 8, 10, 12, 14, 16) $\times$ 0.1. White contours show the naturally weighted integrated$^{12}$CO(3--2) flux with contours of (4, 8, 16, 32, 48) $\times$ 1$\sigma_{\textrm{rms-int}}$ (= 2.0 Jy beam$^{-1}$ km s$^{-1}$) for a beam size of $3.1''$.  
\emph{Top right:} the $^{12}$CO(3--2)/$^{12}$CO(2--1) luminosity ratio with black contours in steps of (4, 8, 12, 16, 20) $\times$ 0.1. White contours show the naturally weighted integrated $^{12}$CO(3--2) flux with contours of (2, 4, 8, 16, 32, 40) $\times$ 1$\sigma_{\textrm{rms-int}}$ (= 2.2 Jy beam$^{-1}$ km s$^{-1}$) for a beam size of $2.5''$.
\emph{Left:} the $^{12}$CO(2--1)/$^{12}$CO(1--0) luminosity ratio with black contours in steps of (3, 4, 5, 6, 8, 10, 12, 14) $\times$ 0.1. White contours show the naturally weighted integrated $^{12}$CO(2--1) flux with contours of (2, 4, 8, 12, 16, 20, 24) $\times$ 1$\sigma_{\textrm{rms-int}}$ (= 0.9 Jy beam$^{-1}$ km s$^{-1}$) for a beam size of $3.1''$. Gray contours show the naturally weighted integrated $^{12}$CO(3--2) flux as in Fig. \ref{HE1029-int-vel-disp}. 
\vspace{0.25cm}}
%
 \label{HE1029-ratio}
\end{minipage}\\
\end{array} $                                                           
\vspace{0.5cm}
\end{figure*}

The velocity ranges from -55 km s$^{-1}$ to 70 km s$^{-1}$ and the velocity gradient at the center is rather perpendicular to the primary bar, consistent with the study of \citet{Krips2007}. These authors find a PA = 90$^\circ$ for the steepest velocity gradient, whereas we find PA = 115$^\circ$ $\pm$ 20$^\circ$,
but with high uncertainty because of the low SNR. According to their simulation, the velocity field can be explained best by a bar potential and the velocity gradient by a bar–driven inflow. 
The velocity dispersion is flat with $\sigma_\textrm{v}$ $\sim$ 50 km s$^{-1}$ at the center, elongated along the 0 km s$^{-1}$-contour (23$^\circ$ vs 24$^\circ$) and the major axis, i.e,. the $^{12}$CO(3--2) transition mainly traces the inner portion of the bulge. 

\begin{figure*}[!htbp]
\centering $
\begin{array}{cc}
\vspace{0.2cm} 
\includegraphics[trim = 0mm 1mm  50mm 1mm, clip, width=0.39\textwidth]{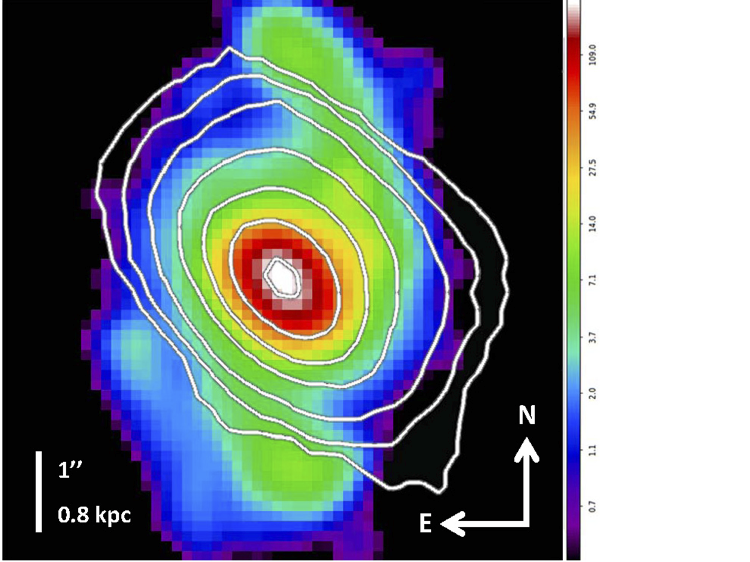}  & 
\hspace{1.3cm}
\includegraphics[trim = 0mm 1mm  50mm 1mm, clip, width=0.39\textwidth]{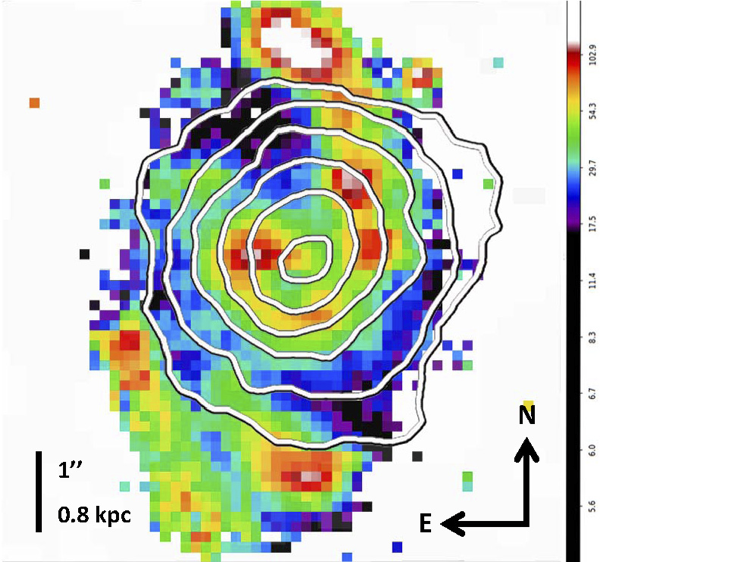}\\ 
\end{array} $                                                           
\caption{HE 1029-1831. 
\emph{Left:} $7'' \times 7''$ SINFONI image \citep{Busch2014} of the integrated flux (arbitrary units) of the narrow Pa$\alpha$ component of the galaxy overlayed with contours of naturally weighted integrated $^{12}$CO(3--2) flux as in Fig. \ref{HE1029-int-vel-disp}. 
\emph{Right:} corresponding equivalent width (arbitrary units) of the narrow Pa$\alpha$ component with contours of uniformly weighted integrated $^{12}$CO(3--2) flux as in Fig. \ref{HE1029-int-vel-disp}.
}
\label{HE1029-overlay}
\end{figure*}


The PV diagrams are consistent with previous measurements \citep{Krips2007}. Along the minor axis (PA = 0$^\circ$), the PV diagram shows
a fast unresolved component with a velocity range of -100 to 100 km s$^{-1}$. Along the major axis (PA = 90$^\circ$),  
the velocities follow a steep slope and a possible x-shape is indicated, with a steep component at offset 0$''$ ranging between -100 to 100 km s$^{-1}$, and a shallower one from -1.5$''$ and -60 km s$^{-1}$ to 1.5$''$ and 60 km s$^{-1}$. Hence, the presence of a kinematically distinct feature in the circumnuclear region cannot be ruled out.

HCO$^+$(4--3) emission was not detected, therefore, we can only give a 3$\sigma$ upper limit for the total flux. With an rms in a 20 km s$^{-1}$ channel of 33 mJy beam$^{-1}$ and the line width of 150 km s$^{-1}$, similar to the $^{13}$CO (2--1) emission line width for the other sources, we obtain 5.3 Jy beam$^{-1}$ km s$^{-1}$ as an upper limit. We estimate a $^{12}$CO(3--2)/HCO$^+$(4--3) luminosity ratio at the peak position of 16 as a lower limit. 

To obtain line ratios for the nuclear region with the PdBI $^{12}$CO (1--0) and (2--1) data of \citet{Krips2007}, we truncated both data sets to a common uv range of 14 - 120 k$\lambda$ (15$''$-1.7$''$), i.e., the uv range of the SMA data so that the spatial scales considered are the same for all data sets.
We used a circular restoring beam size set to the major axis length of the lower transition's beam. The $^{12}$CO (1-0) image was uniformly weighted to achieve a higher resolution, i.e., 3.1$''$ in comparison to natural weighting with 3.5$''$. This might have introduced a bias, but no significant difference between the two approaches are found. The resulting ratio images are shown in Fig. \ref{HE1029-ratio}.
The $^{12}$CO(3--2)/(1--0) line ratio is $\sim$ 1 in the center, 1.4 in the eastern and western regions of emission, and 0.6 in the northern and southern outskirts of emission.
For the $^{12}$CO(3--2)/(2--1) line ratio, we find a value of $\sim$ 2 in the center, 1.6 east and west, and 0.6 north and south of the center. 
As a test we also verified the $^{12}$CO(2--1)/(1--0) ratio at 3.1$'',$ which is still consistent with \citet{Krips2007} when we keep beam smearing in mind.
As a result of the large beam size, we obtained rather average values, which are mainly significant in the central beam area.
The line ratios are discussed in Sect. \ref{sec:12COratios}.

An overlay of the naturally weighted image onto a SINFONI image of the Pa$\alpha$ emission and equivalent width \citep{Busch2014} shows that the warm and dense gas emission is dominated by the circumnuclear star formation ring (Fig. \ref{HE1029-overlay}). The asymmetry to the northwest is caused by star formation/ionized gas clumps along the northern spiral arm. In the uniformly weighted map, these features make the emission peak even shift from the ring center toward them.


\begin{figure*}[!htbp]
\centering $
\begin{array}{cc}
\includegraphics[trim = 1mm 0mm 6mm 2mm, clip, width=0.47\textwidth]{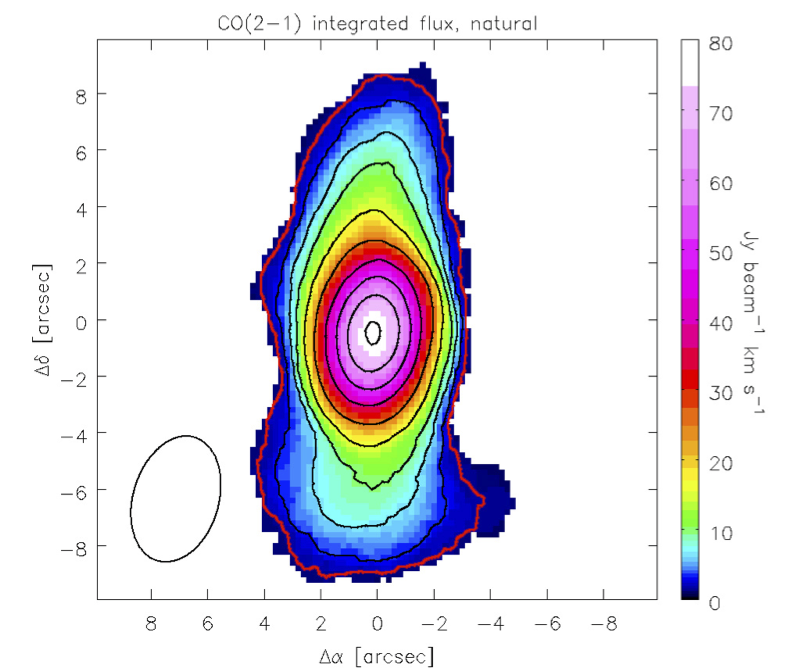}&
\includegraphics[trim = 1mm 0mm 6mm 2mm, clip, width=0.47\textwidth]{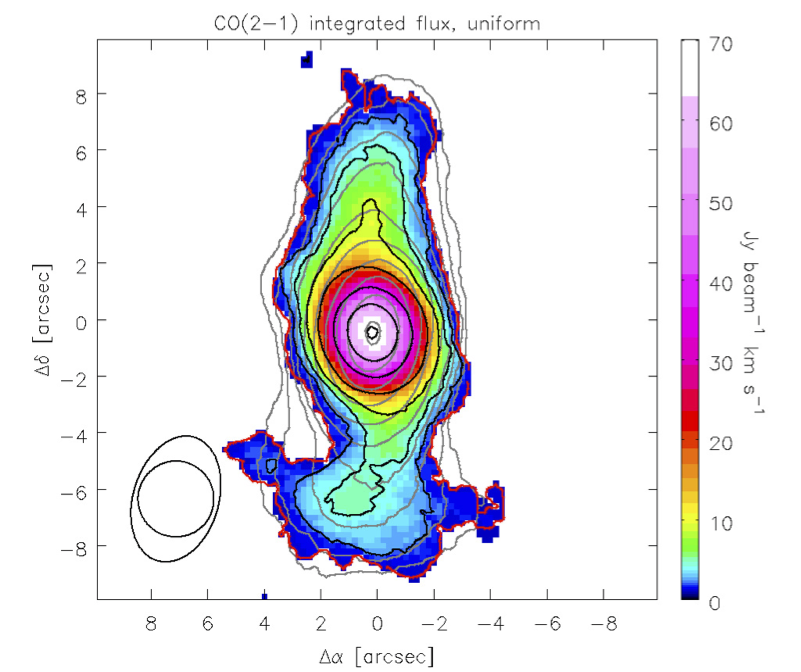}\\
\includegraphics[trim = 1mm 0mm 6mm 2mm, clip, width=0.47\textwidth]{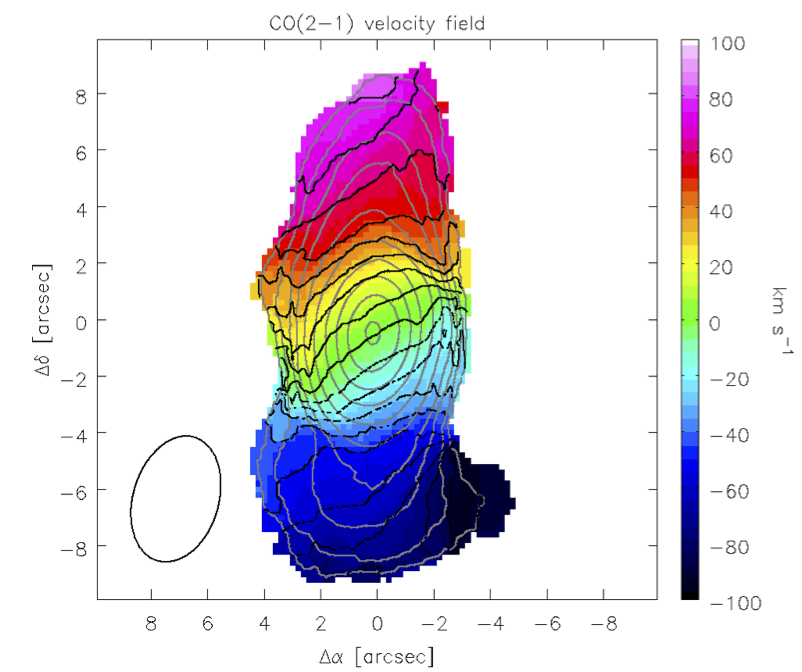}&
\includegraphics[trim = 1mm 0mm 6mm 2mm, clip, width=0.47\textwidth]{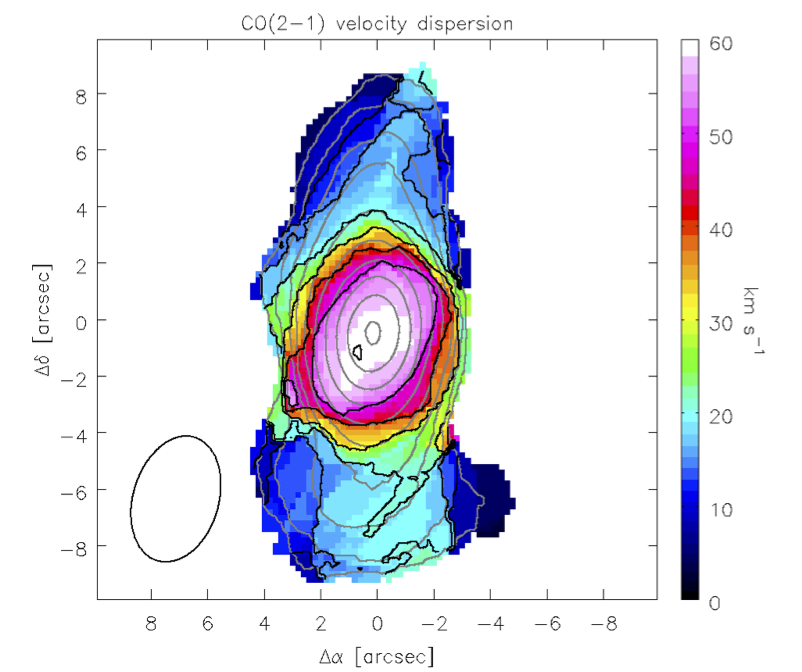}\\
\includegraphics[trim = 1mm 0mm 6mm 2mm, clip, width=0.47\textwidth]{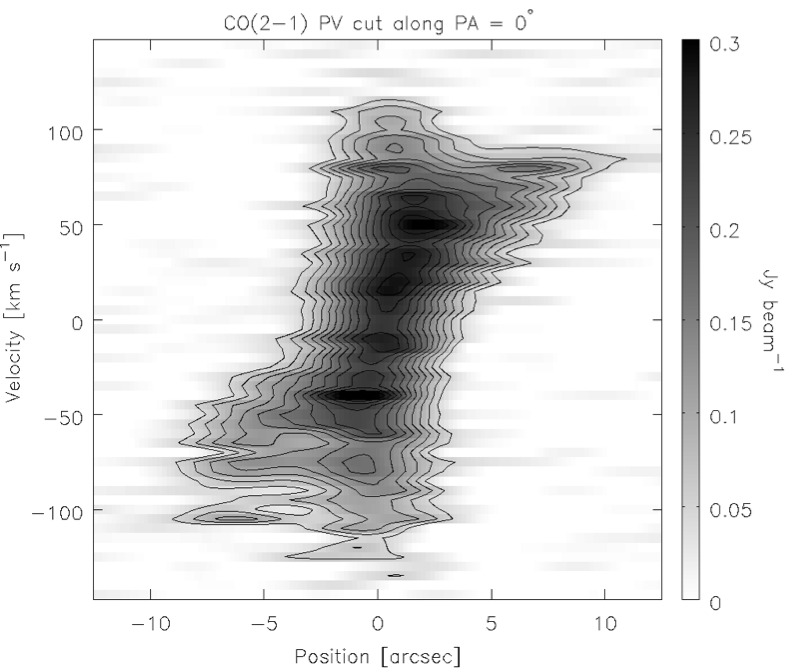}&
\includegraphics[trim = 1mm 0mm 6mm 2mm, clip, width=0.47\textwidth]{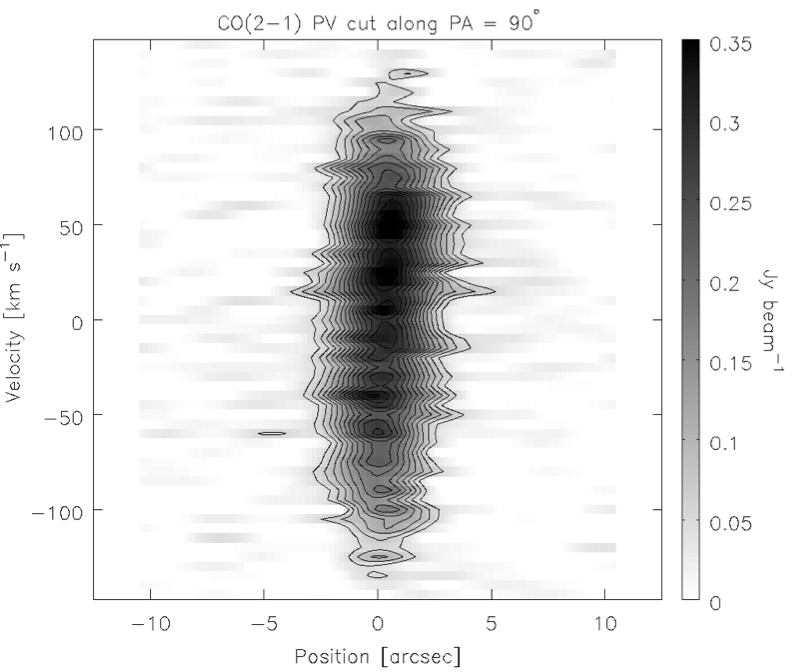}\\
\end{array} $                                                           
\caption{HE 1108-2813. 
\emph{Top panels:} the integrated $^{12}$CO(2--1) flux image for natural weighting (\emph{left}) and uniform weighting (\emph{right}) obtained for a velocity range from -125 to +130 km s$^{-1}$. The beam sizes are $4.5'' \times 3.0''$ and $2.68''$, respectively.
The contours are in steps of (2 (red), 4, 8, 12, 20, 32, 48, 64, 80, 96) $\times$ 1$\sigma_{\textrm{rms-int}}$ (= 0.8 Jy beam$^{-1}$ km s$^{-1}$) and (1 (red), 2, 4, 8, 16, 32, 48, 62) $\times$ 1$\sigma_{\textrm{rms-int}}$ (= 1.1 Jy beam$^{-1}$ km s$^{-1}$), respectively.
\emph{Middle panels:} the corresponding isovelocity (\emph{left}) and velocity dispersion image (\emph{right}) for natural weighting. Black contours are in steps of 10 km s$^{-1}$ from -90 to +80 km s$^{-1}$ and from 10 to 60 km s$^{-1}$ (additional contour at 15 km s$^{-1}$), respectively. Gray contours show the naturally weighted integrated $^{12}$CO(2--1) flux.
\emph{Bottom panels:} the position velocity diagrams along the major (PA = 0$^\circ$, \emph{left}) and minor (PA = 90$^\circ$, \emph{right}) axis.
}
\label{HE1108-int-vel-disp}
\end{figure*}

\subsubsection{HE 1108-2813}
\label{subsubsec:morph-HE1108}

The molecular gas extends along the primary bar and its distribution points to a spiral arm pattern within the bar (Fig. \ref{HE1108-int-vel-disp}). 
The strong nuclear component with PA = -9$^\circ$ $\pm$ 5$^\circ$ cannot have a circular source shape since the beam is oriented at PA = -17$^\circ$ $\pm$ 1$^\circ$, i.e., there must be a noncircular component at a larger PA.
This pattern becomes visible best in the uniform weighting of the uv data. In addition, we see a nuclear component elongated along a PA = 20$^\circ$ $\pm$ 10$^\circ$. 
Toward the tips of the bar the emission increases to secondary maxima and extends at the southern tip into the major spiral arm. 

The velocity field shows a torsion of the isovelocity lines in the central region. 
Similar to the case of HE 0433-1028, we interpret this as an indication of a kinematically different feature in the center, e.g., a nested bar or nuclear spiral/disk. 
The velocity dispersion peaks in the center, as expected for the turbulent motion in a galactic bulge, and has dispersions around $\sigma_\textrm{v}$ $\sim$ 50 - 60 km s$^{-1}$.
The velocity dispersion
shows an elongation along PA = -31$^\circ$ $\pm$ 10$^\circ$.
There is also some enhanced dispersion, i.e., $\sigma_\textrm{v}$ $\sim$ 20 - 25 km s$^{-1}$, observable in the primary bar. These are turbulent regions, i.e., the shock fronts of the leading edges of the bar as already indicated in the total flux image.

\begin{figure}[!htbp]
\centering $
\begin{array}{c}
\includegraphics[trim = 1mm 0mm 6mm 2mm, clip, width=0.47\textwidth]{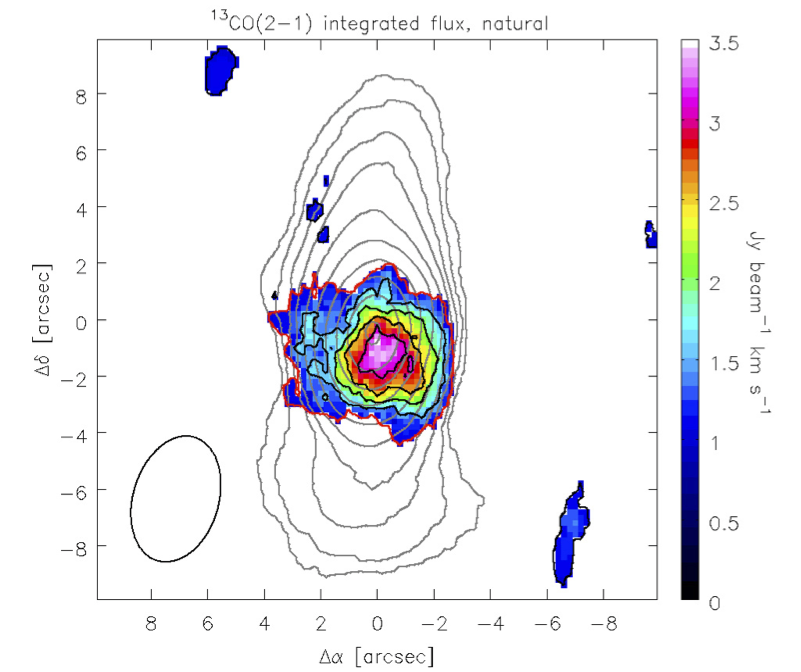}\\
\includegraphics[trim = 1mm 0mm 6mm 2mm, clip, width=0.47\textwidth]{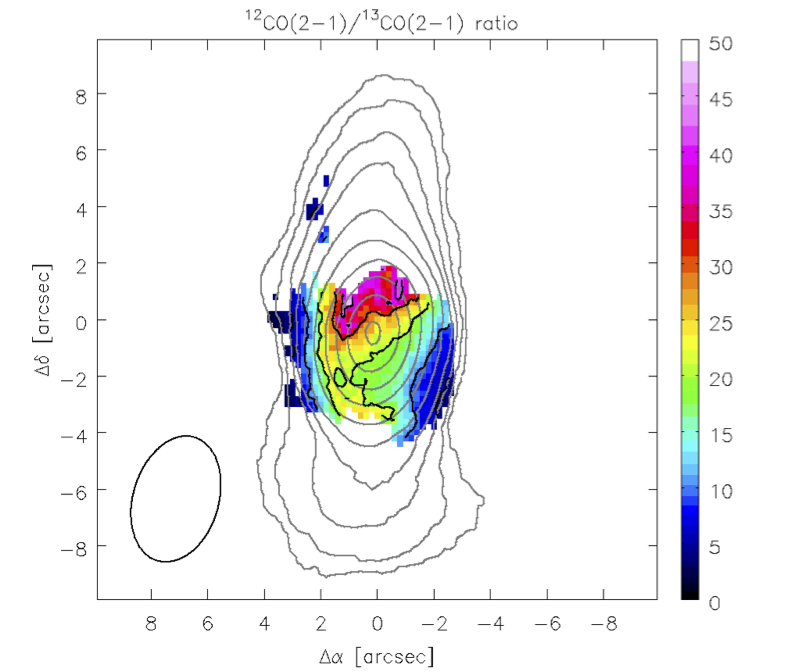}\\
\end{array} $                                                           
\caption{HE 1108-2813. 
\emph{Top:} the integrated $^{13}$CO(2--1) flux image for natural weighting obtained for a velocity range from -75 to 75 km s$^{-1}$. Black contours are in steps of (2 (red), 3, 4, 5, 6) $\times$ 1$\sigma_{\textrm{rms-int}}$ (= 0.5 Jy beam$^{-1}$ km s$^{-1}$). The beam size is $4.5'' \times 3.0''$. 
\emph{Bottom:} the $^{12}$CO(2--1)/$^{13}$CO(2--1) luminosity ratio with black contours in steps of 5, 10, 20, 30, 40.
Gray contours show the naturally weighted integrated $^{12}$CO(2--1) flux as in Fig. \ref{HE1108-int-vel-disp}. 
}
\label{HE1108-ratio}
\end{figure}

\begin{figure}[!htbp]
\centering $
\begin{array}{c}
\vspace{0.4cm}
\includegraphics[trim = 0mm 1mm  50mm 1mm, clip, width=0.39\textwidth]{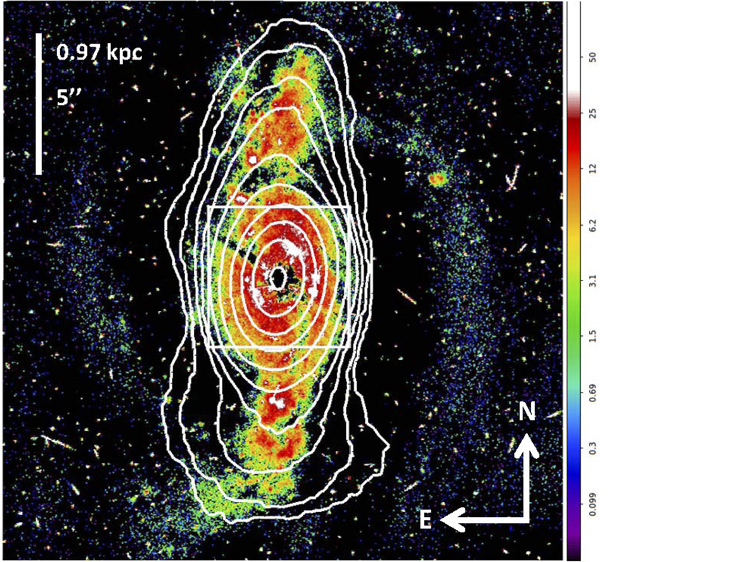}\\
\vspace{0.4cm}            
\includegraphics[trim = 0mm 1mm  50mm 1mm, clip, width=0.39\textwidth]{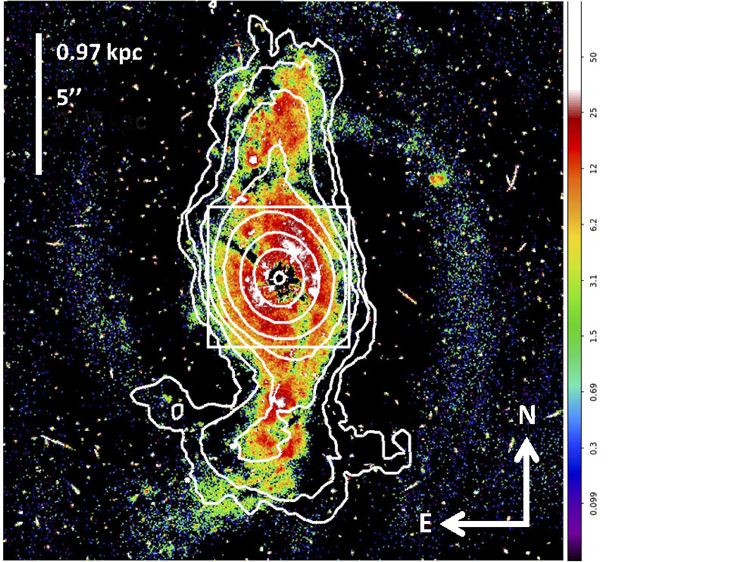}\\
\vspace{0.4cm} 
\includegraphics[trim = 0mm 1mm  50mm 1mm, clip, width=0.39\textwidth]{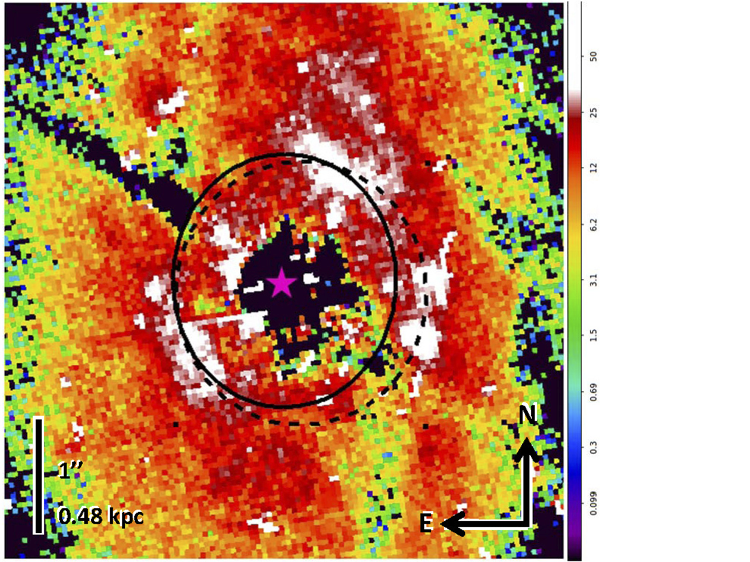}\\
\end{array} $                                                           
\caption{HE 1108-2813. 
\emph{Top:} $20'' \times 20''$ of unsharp-masked HST image (F606, red, HST archive, arbitrary intensity units) of the galaxy overlayed with contours of naturally weighted integrated $^{12}$CO(2--1) flux as in Fig. \ref{HE1108-int-vel-disp}.
The box size is $5'' \times 5''$.
\emph{Middle:} same as top, but with contours of uniformly weighted integrated $^{12}$CO(2--1) flux as in Fig. \ref{HE1108-int-vel-disp}.
\emph{Bottom:} zoomed into the box region of the image. The solid and dashed circles indicate possible circumnuclear ring configurations (see section \ref{subsubsec:morph-HE1108}). The magenta star denotes the AGN's position. 
} 
\label{HE1108-overlay}
\end{figure}     


The distribution of velocities along the major axis clearly shows an x-shape. 
The diagonal part from about -6$''$ to 7$''$ and -100 to 80 km s$^{-1}$ can be associated with the rotation curve of a disk. The vertical part around 0$''$ offset and -125 to 120 km s$^{-1}$ represents an unresolved and kinematically decoupled nuclear component. The minor axis PV cut strongly resembles the vertical component along the major axis, hinting at the same origin.
The (projected) velocities in the circumnuclear region exceed the disk velocities. However, flux in the outskirts of the bar/disk might not have been detected for sensitivity reasons and the nuclear component might have an inclination that is different from the bar/disk.

The $^{13}$CO emission (Fig. \ref{HE1108-ratio}) is tentatively detected in the nuclear region and peaks slightly off-center by $\sim$ 0.7$''$ (i.e., $<$ 0.23 $\times ~\theta_\textrm{minor}$) toward the southwest. 
The $^{12}$CO/$^{13}$CO line ratio is 20 at the $^{13}$CO peak, 25 at the $^{12}$CO peak, and goes down to 5 toward the outskirts of the $^{13}$CO emission, i.e., r $\sim$ 2$''$, especially east of the center. The ratio distribution looks similar to the one of HE 0433-1028.

The AGN in the HST F606W image \citep{Malkan1998} is strong enough to outshine the nuclear region and saturate the point spread function (PSF). We removed a scaled and truncated PSF, which was generated for the pixel positions of the AGN region by the Tiny Tim software \citep{KristHook2011}, to improve
the image fidelity. We adjusted the flux of the PSF according to the difference of the flux levels of the diffraction spikes and the background and clipped the peak at a value corresponding to the difference of the maximum counts and the background emission of the circumnuclear region. With this approach, we sufficiently reduced the diffraction spikes. However, there is obviously an asymmetry in the PSF visible as negative spikes along the diagonals of the image. 

The unsharp masked version of this image in Fig. \ref{HE1108-overlay} reveals manifold details on the morphology. The primary bar appears as clumpy bright two-armed spiral containing luminous star formation regions. Embedded inside are dust lanes that wind off-center toward the nuclear region and connect to it. These shock zones at the bar leading edges are also outlined by several star formation spots. The circumnuclear region contains bright star-forming regions appearing to form a ring-like structure that could most likely be the inner Lindblad resonance. 

We find two possible ring shapes indicated by the solid and dashed lines in Fig. \ref{HE1108-overlay}, which match the constellation of the star-forming region.
The solid circle has a size of 2.0$''$ $\times$ 2.2$''$, i.e., 0.97 kpc $\times$ 1.07 kpc, and is centered on the AGN. The dashed circle is marginally larger (2.2$''$ $\times$ 2.4$''$, i.e., 1.07 kpc $\times$ 1.16 kpc) and slightly offset from the AGN by $\sim$ 100 pc, i.e., $\Delta \alpha$ = 0.14$''$, $\Delta \delta$ = -0.14$''$. The diameter of about 1 kpc corresponds well to the minor axis $d_\mathrm{FHWM}= $ 0.8 kpc of the fitted Gaussian component. 
Because of the proximity of the eastern and western star-forming clusters to the spiral arms and the devoid region in the south, the star formation regions seem to follow the "string of pearls" model. This model proposes that the clusters form in a very short-lived starburst when the gas passes the overdensity region in the ring. These are typically located around the connection points, i.e., $x_1$--$x_2$ - orbit intersection region, where the gas streams from the bar spiral arm to the circumnuclear ring \citep[see, e.g., Fig. 15 in][]{Regan2003}. 
However, to finally confirm or rule out this model, the age gradient of the cluster populations along the ring needs to be investigated.
The clusters are expected to be youngest at the overdensities, and then sequentially older along the inflow direction from the spiral arm \citep[see, e.g.,][]{Boker2008}.

As an alternative scenario, the two bright star formation regions in the ring could not perfectly mark the aligned tips of a possible nuclear bar at a PA = -37$^\circ$ $\pm$ 10$^\circ$, corresponding to the central velocity dispersion field. 
Further inward from the circumnuclear ring, the HST image is not reliable because of the problematic PSF subtraction.

An overlay with the $^{12}$CO emission shows that the dust lanes correspond well to the $^{12}$CO spiral arm pattern within the bar and that the $^{12}$CO emission extends even into the southern galactic spiral arm. Overlaying the uniformly weighted map outlines the dust lanes even better. 
The overdensities nearby/around the intersection region become visible in, e.g., the $^{13}$CO(2--1) emission, where the gas is cold and dense enough. 
In the central region, the ISM might be too hot and/or dense to excite the $^{13}$CO(2--1) transition \citep[compare][]{Huttemeister2000}.

\subsubsection{A note on the kinematics}
\label{sec:kin}

All three galaxies show signs of an unresolved kinematically distinct component in the center. This is not unusual for Seyfert galaxies; e.g., \citet{Dumas2007} find, in a sample of matched active-inactive galaxy pairs, that the gas velocity fields at small radii (r < 500 pc) are more disturbed in Seyfert than in inactive galaxies. The deviations from axial symmetry range from wiggles along the gas kinematic minor axis to a high misalignment of the kinematic major axes of gas and stars where the latter is only found in Seyfert galaxies. This indicates a close relation between the central kinematics and the feeding of the SMBH. 
Similarly, \citet[][]{Falcon-Barroso2006} and \citet[][]{Hicks2013} find central drops in gas/stellar velocity dispersion to occur with intense star formation in the same place and a high concentration of gas (r $\lesssim$ 500 pc). 
All these features can be interpreted as a dynamically cold (compared to the bulge) nuclear structure (e.g., a disk) and appear to be crucial for the star formation and AGN activity. 
%

\subsection{Dynamical mass}
\label{sec:Mdyn}

We estimate the dynamical masses following \citet[][]{Lequeux1983}, i.e.,
\begin{equation*}
M_{\textrm{dyn}}(r<R)=2.325 \times 10^5 \; \alpha_M \; R \; \left(\dfrac{v(R)}{\sin{i}}\right)^2,
\end{equation*}
with the radius $R$ of the enclosed mass in kpc, the projected velocity $v(R)$ in km s$^{-1}$, $i$ the inclination of the galaxy, and $M_{\textrm{dyn}}$ in $M_\odot$. The factor $\alpha_M$ depends on the disk geometry model, 1 for spheroidal, 0.6 for flat. We adopted an intermediate value of 0.8. The assessed distances and corresponding velocities from the PV diagrams, inclinations and masses are given in Table \ref{Mdyn}. The inclinations are calculated from the major and minor axis diameters according to \citet{Hubble1926} \citep[see also][]{vdBergh1988}. 
We obtain dynamical masses of $(1.5 - 6.7) \times 10^9 M_\odot$.

The ULIRG conversion factor derived gas masses make up 10 - 50\% of the dynamical mass, and for Galactic conversion the gas mass exceeds the dynamical mass by a factor of 0.8 - 3.0.

In fact, the dynamical masses derived from our data are likely to be underestimated or not accurate. Firstly, except for HE 1108-2813, the data gives no hints on whether the turn-over point from solid body rotation to flat rotation has been detected or not. It is very likely that a significant amount of emission at higher velocities, i.e., outside the FWZI range, and larger distances is below a $1 - 2 \sigma$ level. While the effect on the $^{12}$CO luminosity is marginal, it makes a considerable difference in the dynamical mass. Uncertainties of, e.g., 25\% in the distance and 13\% in the velocity result in a $\sim$ 30\%-error for the dynamical mass (HE 1029-1831). Secondly, the inclination is based on the assumption that the visible luminosity distribution, which the major and minor axes are obtained from, is perfectly circular. Therefore, our used inclinations are not exact. For comparison we also derive inclinations necessary for the gas masses to correspond to 10\% \citep{Young1991} of the dynamical masses. 
For a ULIRG conversion factor, we find inclinations of 21 - 38$^\circ$, and for the Galactic conversion factor, the galaxies are virtually face on (see Table \ref{Mdyn}). 

If we assume the dynamical mass to be on the order of the bulge mass, we can use the black hole mass - bulge mass correlation to estimate the dynamical masses. The masses of the black holes are on the order of $M_\mathrm{BH} \sim 10^{7-8} M_\odot$, therefore the dynamical/bulge masses must be on the order of $M_\mathrm{dyn} \sim (0.3 - 5.6) \times 10^{10} M_\odot$ \citep{Sani2011,Kormendy2013}. The bulge masses for HE 1029-1831 and HE 0433-1028 obtained from these relations (Table \ref{Mdyn}) are 2.5 and 5 times larger than the gas dynamical values, while the 30\% lower value for HE 1108-2813 is still within the error range of the gas dynamical mass. Recently,
\citet{Busch2014} obtained a dynamical mass of $M_\mathrm{dyn} = 6 \times 10^{9} M_\odot$ for HE 1029-1831 based on stellar velocity dispersion. This value is four times larger than that yielded from our data. However, the molecular gas still makes up more than 43 \% of the bulge masses in the case of a Galactic conversion factor $\alpha_\mathrm{MW}$. This may imply that the assumption of self-gravitating giant molecular clouds (GMCs) does not hold for these galaxies (see Sect. \ref{sec:ISM}).

\begin{table*}[!htb]
\centering
\caption{Dynamical masses}
\begin{tabular*}{\textwidth}{@{\extracolsep{\fill}} ccccccccc}
\toprule
Object          & \multicolumn{2}{c}{Radius $R$}        & $v$           & $i$\tablefootmark{1}    & $M_\textrm{dyn}$      & $M_\textrm{mol}/M_\textrm{dyn}$\tablefootmark{2}      & $i_{M_\textrm{mol}=10\%M_\textrm{dyn}}$\tablefootmark{2}        &       $M_\textrm{dyn (bulge)}$\tablefootmark{3}      \\                                                                                      &       [$''$]  &       [kpc]   & [km/s]        & [$^{\circ}$]            & [$10^9 M_\odot$]      &                                       & [$^{\circ}$]                                    &       [$10^9 M_\odot$]        \\
\midrule  
%
%
%
%
%
%
HE 0433-1028    & 1.0 $\pm$ 0.3 & 0.7 $\pm$ 0.2 &               100  $\pm$ 10      & 42            & 3.0 $\pm$ 0.8         & 0.3 - 2.1 & 21 - $\;\,$8      &               14.5 \\
HE 1029-1831    & 1.0 $\pm$ 0.3 & 0.8 $\pm$ 0.2 &  $\;\,$80  $\pm$ 10   & 53              & 1.5 $\pm$ 0.4         & 0.5 - 3.0 & 21 - $\;\,$8      &  $\;\,$3.7 \\
HE 1108-2813    & 6.0 $\pm$ 1.0 & 2.9 $\pm$ 0.5 &  $\;\,$80  $\pm$ 20   & 46              & 6.7 $\pm$ 2.0         & 0.1 - 0.8 & 38 -      15              &  $\;\,$4.7 \\
\bottomrule

\end{tabular*}
\tablefoot{
\tablefoottext{1}{Inclination based on major and minor axis values taken from NASA/IPAC Extragalactic Database (NED),}
\tablefoottext{2}{range given by ULIRG and Galactic mass conversion factor,}
\tablefoottext{3}{bulge mass; average of the results of the $M_\textrm{BH}$-$M_\textrm{bulge}$ relations given in \citet{Sani2011} and \citet{Kormendy2013}, derived from the average of the black hole mass range limits given in Table \ref{lit_props}.}}
\label{Mdyn}
\end{table*}

\begin{table*}[!htb]
\centering
\caption{Dust properties derived from the MIR/FIR fit}
\begin{tabular*}{\textwidth}{@{\extracolsep{\fill}} cccccccccccc}
\toprule
Object  & $S_{3\sigma \textrm{cont}}$   & $L_\textrm{IR}$       & $L_\textrm{FIR}$      & $T_\textrm{fit}$ & $T_\textrm{peak}$    & $\beta$       & $M_\textrm{dust}$         & $M_\textrm{mol}/M_\textrm{dust}$\tablefootmark{1} \\  
        & [mJy beam$^{-1}$]             & [$10^{11} L_\odot$]   & [$10^{11} L_\odot$]       & [K]               & [K]               &               & [$10^6 M_\odot$]        & [$10^2$]                         \\   
\midrule 
%
%
%
%
%
%
HE 0433-1028  & $\;\,$4.1 (220 ~GHz) & 2.2 $\pm$ 0.1  & 1.1 $\pm$ 0.1   & 58 $\pm$ 1  & 38 $\pm$ 1  & 1.5 $\pm$ 0.1 &  4.4 $\pm$ 0.6  & 2.3 - 14.1 \\                                                      
HE 1029-1831  &      11.1 (330 ~GHz) & 2.0 $\pm$ 0.3  & 1.3 $\pm$ 0.2   & 58 $\pm$ 3  & 36 $\pm$ 3  & 1.7 $\pm$ 0.1 &  4.2 $\pm$ 0.6  & 1.8 - 10.7\\                                                         
HE 1108-2813  & $\;\,$2.7 (220 ~GHz) & 0.9 $\pm$ 0.1  & 0.6 $\pm$ 0.1   & 56 $\pm$ 2  & 35 $\pm$ 2  & 1.9 $\pm$ 0.1 &  1.6 $\pm$ 0.2  & 5.6 - 33.5 \\                                                      
%
\bottomrule

\end{tabular*}
\tablefoot{
\tablefoottext{1}{Range given by ULIRG and Galactic mass conversion factor.}}

\label{dust}
\end{table*}

\begin{table*}[!htb]
\centering
\caption{SFRs and surface densities}
\begin{tabular*}{\textwidth}{@{\extracolsep{\fill}} cccccc}
\toprule
                & $SFR_{\textrm{IR}}$   & $SFR_{\textrm{1.4 GHz}}$      & $\tau_{\textrm{1.4 GHz-SFR}}$    & $\Sigma_{mol}$                       & $\Sigma_\textrm{SFR}$    \\     
Object          &                       &                               & H$_2$+He                                 & \multicolumn{2}{c}{within 3 $\times$ $d_\textrm{FWHM 1,2}$}   \\     
                & \multicolumn{2}{c}{[$M_\odot$~yr$^{-1}$]}             & [Gyr]                            & [$M_\odot$~pc$^{-2}$]                        & [$M_\odot$~kpc$^{-2}$yr$^{-1}$] \\      
\midrule                                                                                                                                                                                  
%
%
%
%
%
%
HE 0433-1028    & 33                    &       32                      & 0.03 - 0.19                      & 47 - 280                             & 1.4                              \\     
HE 1029-1831    & 31                    &       25                      & 0.03 - 0.18                      & 92 - 553                             & 3.1                              \\     
HE 1108-2813    & 14                    &       13                      & 0.07 - 0.42                      & 81 - 485                             & 1.1                              \\     
\bottomrule
\end{tabular*}
\label{SFR}
\end{table*}

\begin{figure}[!!htb]                                                     
\centering $                                                            
\begin{array}{c}  
\vspace{0.2cm}
\includegraphics[trim = 15mm 15mm 19mm 33mm, clip, width=0.47\textwidth]{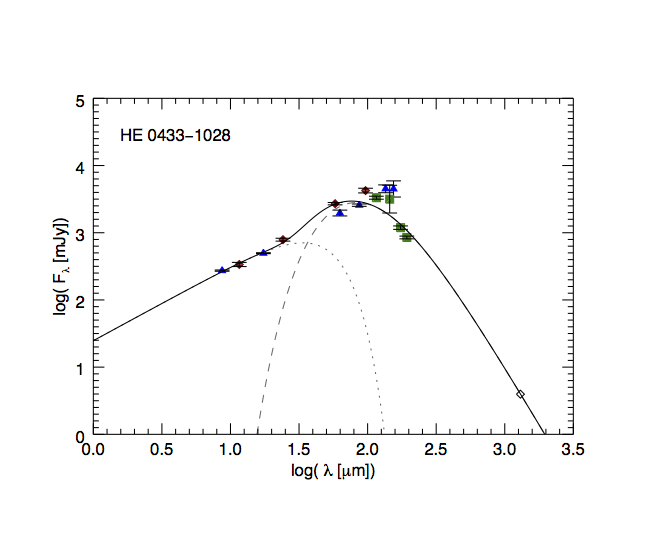}\\
\vspace{0.2cm}
\includegraphics[trim = 15mm 15mm 19mm 33mm, clip, width=0.47\textwidth]{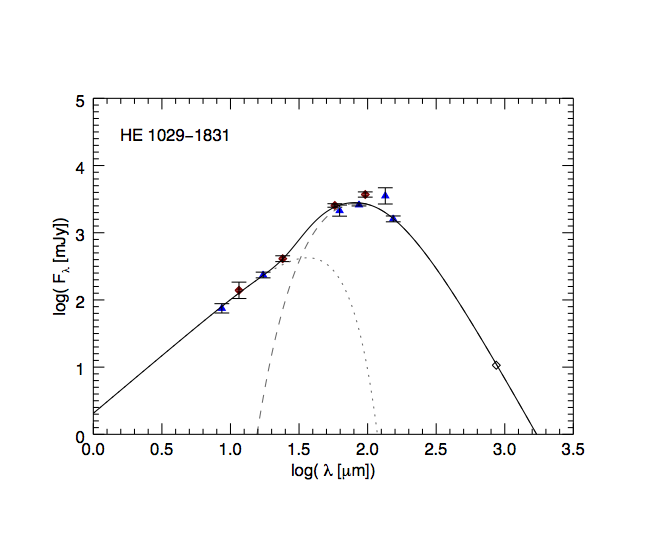}\\
\vspace{0.2cm}
\includegraphics[trim = 15mm 15mm 19mm 33mm, clip, width=0.47\textwidth]{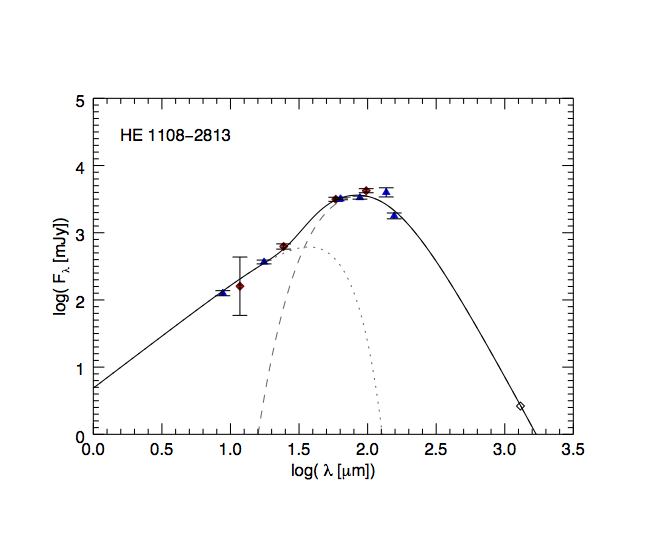}
\end{array} $
\caption{SED fits to the IRAS (red diamonds), AKARI (blue triangles), and ISO data (green squares, only for HE 0433-1028 available) for each source to constrain the properties of the dominating dust component. This SED fitting technique described by \citet{Casey2012} uses a single dust temperature graybody component (gray dashed line) and a MIR power-law component (gray dotted line) to account for the warmer dust emission, i.e., a superposition of several graybodies at different temperatures. The resulting fit curve is depicted as solid black line. The white diamonds represent the SMA upper flux limits, which have been used in the fit.} 
\label{SED-fits}
\end{figure}

\subsection{Continuum emission and dust properties}
\label{sec:dust-prop}

No continuum emission at 220 GHz and 330 GHz, respectively, was detected. Therefore, we give $3\sigma$ upper limits determined from the line-free channels, i.e., excluding the range of $-200$ km s$^{-1}$ - $+200$ km s$^{-1}$ around the emission lines (see Table \ref{dust}).

To estimate dust masses and temperatures, we used the far-IR (FIR) spectral energy distribution (SED) fitting technique of \citet{Casey2012}. Publicly available as IDL code, this method fits a combination of a modified single-temperature graybody component for the FIR ($>$ 50 $\mu$m) and a power-law component for the mid-IR (MIR) ($<$ 50 $\mu$m) to data sets with at least three FIR photometric measurements. 
The first component emerges from cold dust heated by the starburst emission of the entire galaxy. The second component accounts for hot dust heated by an AGN, or dense, hot starburst regions, and can be envisaged as the sum of several graybodies at different temperatures.

The data available for each galaxy comprise six AKARI and four IRAS measurements, and in the case of HE 0433-1028, additional four ISO measurements. We used the SMA continuum upper limits as sub-mm data points to give the sub-mm slope a rough lower limit. The fits are shown in Figure \ref{SED-fits} and the results are listed in Table \ref{dust}.


From the fits, we obtain IR (8-1000 $\mu$m) luminosities on the order of $10^{11}$ $L_\odot$ classifying two of the galaxies as LIRGs. HE 1108-2813 is slightly below the LIRG limit. Luminosities based on the IRAS fluxes (Table \ref{lit_props}) are larger by a factor of 1.3. The FIR luminosities from both methods are consistent when omitting the extrapolation factor for the IRAS based method \citep{Sanders1996}. This is not surprising since the IRAS based method estimates the luminosity via a template of a sample averaged SED, whereas the SED fit takes individual slopes of the SED on both sides of the peak into account. 
The intrinsic dust temperatures of the galaxies are in the range of $T_\textrm{fit}$ = 55 - 59 K. From this to be distinguished are the peak dust temperatures $T_\textrm{peak}$ = 35 - 38 K obtained from Wien's displacement law under the assumption of a single blackbody. 
Since this temperature is determined by the peak wavelength that is more or less covered by the IRAS and ISO fluxes, it does not differ much between fitting techniques and is comparable to results from established methods of dust temperature estimation.
The dust emissivity $\beta$ covers typical values of $1.5 - 1.9$ \citep[][and references therein]{Casey2012}.

Extrapolating the flux to 850 $\mu$m and assuming a dust absorption coefficient of $\kappa$ = 0.15 m$^2$ kg$^{-1}$ \citep{Weingartner2001}, we obtain upper limits for the dust masses of $M_{\textrm{dust}}$ = (1.6 - 4.4) $\times$ $10^6$ $M_{\sun}$.
Fitting without the sub-mm data gives even lower dust masses.
%
%
The ratio between our gas and dust masses corresponds to 1.8 - 5.6 times the standard ratio of 100 for ULIRG typical luminosity-mass conversion and 11 - 34 for Galactic conversion.
Typically, (U)LIRGs 
are found to have a gas-to-dust mass ratio of 200 - 350 \citep{Sanders1991,Contini2003,Seaquist2004,Wilson2008}. This implies that the molecular gas in these three galaxies might be better characterized by a ULIRG typical luminosity-mass conversion and a ULIRG typical gas-to-dust mass ratio. 

However, since only one data point (upper limit) longward of $\sim$ 200 $\mu$m is available, the dust mass estimates comprise several uncertainties. 
Moreover, this sub-mm interferometric data point only traces emission over a limited range of spatial scales, whereas the MIR/FIR data trace the total flux. 
For example, \citet{Wilson2008} found, for a sample of galaxies at similar redshift range and spatial resolution, that $\gtrsim 50 \%$ of the continuum flux can be missed by the interferometer.
Obviously, a significant fraction of the sub-mm dust emission extends to moderately large spatial scales. Doubling our sub-mm fluxes yields dust masses that are roughly twice as large.
Apart from the unknown emissivity, it is unclear whether there is only a single dominating cold blackbody component and whether it is traced reliably by the MIR/FIR fluxes. It is possible that a significant amount of FIR-to-sub-mm flux has been missed or overestimated in the interpolation. 
In addition, the dust mass also depends on the dust absorption coefficient applied. 

\section{Star formation properties}
\label{sec:SF}

In order to assess the star formation activity, we calculate the star formation rate (SFR) from $L_\textrm{IR}$ (Table \ref{dust}) using the calibration of 3.99 $\times ~10^{-37}~M_\odot$ yr$^{-1}$ W$^{-1}$ from \citet[][]{Panuzzo2003}. We obtain $SFR_{\textrm{IR}} ~\sim$ 30 $M_\odot$~yr$^{-1}$ for the two more IR-luminous galaxies and $SFR_{\textrm{IR}}~=$ 14 $M_\odot$~yr$^{-1}$ for HE 1108-2813 (see Table \ref{SFR}). 

Tracing the $SFR$ via the FIR emission assumes that young stars dominate the heating of interstellar dust and that the dust is optically thick. This only holds for starburst galaxies and dusty nuclear starbursts \citep{Panuzzo2003}. In normal galaxies, not all UV/visible emission from the young stars is absorbed by the dust and a cooler diffuse dust component adds to the FIR spectrum. This IR cirrus emission is attributed to more extended dust heated by the general stellar radiation field which, in the case of early-type galaxies, constrains a significant contribution of visible radiation from older stellar populations \citep{Kennicutt1998}. In addition, FIR emission from an AGN can also skew the FIR-SFR relation.

We also calculate the radio-SFR based on the 1.4 GHz fluxes listed on the NASA/IPAC Extragalactic Database (NED) and using the calibration of 6.35 $\times ~10^{-22}~M_\odot$ yr$^{-1}$ W$^{-1}$ Hz from \citet{Murphy2011}. Again a certain AGN contribution to the flux cannot be ruled out \citep{Serjeant2002}. As a result, the $SFR_{\textrm{1.4 GHz}}$ are consistent with the $SFR_{\textrm{IR}}$ within 1 $M_\odot$~yr$^{-1}$ except for the one for HE 1029-1831, which is 6 $M_\odot$~yr$^{-1}$ lower than $SFR_{\textrm{IR}}$. 
Busch et al. (private communication) obtained a SFR of $\sim ~7~ M_\odot$yr$^{-1}$ for HE 1029-1831 based on Br$\gamma$ emission within a radius $r <$ 1$''$. This value is still compatible with ours when we consider the different mean ages of the stellar populations that are contributing to the emission in the wavelength regime probed. While ionized hydrogen emission is thought to be dominated by stars younger than 10 Myr on average, the FIR emission traces age ranges up to 100 Myr \citep[][]{Kennicutt2012}.
However, the SFR seems to be on the order of a few 10 $M_\odot$~yr$^{-1}$ for all three sources. 

Given the $SFR_{\textrm{1.4 GHz}}$, their current molecular gas reservoirs last for 30 - 420 Myr, depending on the conversion factor and neglecting gas recycling. As a current mass reservoir, we used the values obtained from our SMA observation. Using masses based on the IRAM 30m and BIMA fluxes \citep{Bertram2007,Krips2007}, our adopted cosmology and mass conversion factors, we find 70 - 490 Myr for each of the three galaxies.

In considering the mass and $SFR_{\textrm{1.4 GHz}}$ per unit area, we have no information other than the sizes of the Gaussian components fitted to the central component of the $^{12}$CO emission. As a conservative upper limit, we use 3 $\times$ $d_\textrm{FWHM 1,2}$ (Table \ref{CO-source-fit}) of the $^{12}$CO Gaussian for the size of the $^{12}$CO and the FIR emission. This yields a molecular gas mass surface density of $\sim$ 50 - 550 $M_\odot$~pc$^{-2}$, regarding the fraction of gas contained in this Gaussian (Table \ref{CO-source-fit}) and a SFR surface density of 1.1 - 3.1 $M_\odot$~kpc$^{-2}$yr$^{-1}$. Assuming a size of only one $d_\textrm{FWHM 1,2}$ in diameter, the values increase by a factor of 9. 
The above mentioned Br$\gamma$-SFR for HE 1029-1831 corresponds to a SFR surface density of 3.6 $M_\odot$~kpc$^{-2}$yr$^{-1}$ matching out result.

According the overview given in \citet{Kennicutt1998}, the central regions of our galaxies fulfill the criterions of a circumnuclear starburst with SFR time scales $\tau_{\textrm{SFR}} \lesssim $ 1 Gyr, a gas surface density $\Sigma_\textrm{mol} \gtrsim $ 100 $M_\odot$~pc$^{-2}$ , and a SFR surface density $\Sigma_\textrm{SFR} \gtrsim $ 1 $M_\odot$~kpc$^{-2}$yr$^{-1}$.

Indeed, rescaled to a surface diameter of 2 $\times$ $d_\textrm{FWHM 1,2}$ as a compromise between the size extremes and to the conversion factors of the comparison samples, our three galaxies lie in the middle of the starburst sample of \citet{Kennicutt1998a} and in a transition region between the star-forming galaxies (SFG) and sub-millimeter galaxies (SMG) by \citet{Genzel2010} at z = 1 - 2.5 and z = 1 - 3.5, respectively. 

The calculated depletion times in Table \ref{SFR} are based on the assumption of the same surface for molecular gas and star-forming region. 
This yields log $\tau_{\textrm{SFR}}$ = 8.1 - 8.4 for $\alpha_\textrm{CO} =$ 3.2 $M_\odot$~(K km s$^{-1}$ pc$^2$)$^{-1}$.
With $\log sSFR $ = - (8.8 - 8.1) ($sSFR = SFR/M_\star$) assuming the stellar mass $M_\star$ to correspond to 90 \% $M_\textrm{dyn}$ for HE 1108-2813 and 90 \% $M_\textrm{dyn (bulge)}$ for the other two sources (see Table \ref{Mdyn}), this puts them in the middle of the transition region from LIRG to ULIRG and SFG (z $>$ 1) to SMG (z $>$ 2) in \citet[][Fig. 9]{Saintonge2011}.
Using an $\alpha_\textrm{CO ULIRG} =$ 1 $M_\odot$~(K km s$^{-1}$ pc$^2$)$^{-1}$ as in \citet[][Fig. 9]{Saintonge2011} shifts them 0.5 dex down, away from the high-z SFG and right into the ULIRG and SMG regime.

Nevertheless, the extreme case of the surface diameter for one quantity to be as much as three times the surface diameter of the other (i.e., gas mass and SFR) results in a maximum variation of the depletion time of $\sim$ 1 dex. 
Within this range theses quantities
cover the local/distant normal star-forming galaxies regime with $\tau_{\textrm{SFR}} \gtrsim $ 1 Gyr as well as the (U)LIRG/starburst regime with $\tau_{\textrm{SFR}} \lesssim $ 0.1 Gyr. 

Since these three galaxies belong to the most luminous of our CO-tested subsample, they are not necessarily representative of the whole sample. 
In fact, we have shown in \citet{Bertram2007}, via a $L_\textrm{FIR}$ vs. $L'_\textrm{CO}$ plot, that the LLQSOs from our CO-tested subsample occupy the transition region from Seyfert galaxies to LIRGs/QSOs rather than the regime of nearby galaxies. The latter corresponds to samples such as that found in \citet[]{Bigiel2008}. The LLQSOs appear to be shifted by 1 dex to higher $L_\textrm{FIR}$ compared to the nearby galaxies. Consequently, LLQSOs and nearby galaxies cannot be described by the same linear relation.
In \citet[][comparison of LLQSOs, referred to as HE sources, NUclei of GAlaxies (NUGA), and Palomar-Green QSO sources]{Moser2012} one can see that $M_\textrm{H2}$ is distributed around similar values for all three samples, while $L_\textrm{IR}$ and $L_\textrm{FIR}$ show a much steeper slope with redshift. Obviously ,the star formation efficiency rises with redshift for these three samples. 

%
%
%
%

\section{Properties of the interstellar matter}
\label{sec:ISM}

All three galaxies show a $^{12}$CO(2--1)/(1--0) ratio of $r_{21} \sim 0.5$. The $^{12}$CO/$^{13}$CO(2--1) ratio, measured for two of the galaxies, is $R_{21} \gtrsim 20$ on the nucleus. For the other galaxy, we observed the $^{12}$CO(3--2)/(1--0) and $^{12}$CO(3--2)/(2--1) ratio with values of $r_{31} \sim 1$ and $r_{32} \sim 2$. For the $^{12}$CO(3--2)/HCO$^+$(4--3) ratio in the same galaxy, we can give a lower limit of $16$.

\subsection{$^{12}$CO/$^{13}$CO(2--1) ratio}
\label{sec:1213CO}

The $^{12}$CO/$^{13}$CO(1--0) ratio in the Milky Way is $R_{10} \sim 5-8$. The lower value is typical for cool 
GMCs ($T_\textrm{k} \leq 20$ K) and increases toward the center of our galaxy \citep[][and ref. therein]{Solomon1979,Polk1988,Oka1998}.
The ratio reaches $R_{10}, R_{21} \sim 10$ in the centers of local spirals \citep[][and ref. therein]{Sawada2001,Meier2000,Mao2000,Sakamoto2006b,Sakamoto2007} and nearby bright IR galaxies \citep[][]{Tan2011}. An $R_{10} \sim 10-15$ is typical in inner regions of normal starburst galaxies \citep[][]{Aalto2007}. Ratios of $R_{10} > 20$ have only been observed in a few luminous mergers \citep[e.g.][]{Aalto1991b,Casoli1992,Aalto2010,Costagliola2011},
with extreme ratios of $R_{10} \sim 40$ \citep[e.g.,][]{Huttemeister2000}. The ratios of $^{12}$CO/$^{13}$CO in the (1--0) and (2--1) transition behave similarly, depending on the excitation state \citep[][]{Papadopoulos2012a}. 
The change in the line ratio from Milky Way to mergers reflects a crucial change in the ISM properties from a quiescent GMC typical gas state to a turbulent hot gas state. 
 
In order to yield a high line ratio, the optical depth $\tau$ of the $^{12}$CO line emission needs to be moderate or at least significantly larger than the optical depth of the $^{13}$CO line emission. 
One efficient way to achieve this is a temperature gradient.
Since the intensity of the $^{13}$CO line is a steeper function of the excitation temperature than the $^{12}$CO intensity, 
$\tau_{^{12}\textrm{CO}} / \tau_{^{13}\textrm{CO}}$, and also, $R_{21}$, increase with temperature \citep[i.e., radiative trapping;][]{Wilson1999,Goto2003}.
%

Another important impact on the opacity is the gas dynamics. In diffuse, gravitationally unbound molecular gas, produced by cloud collisions in the deep potential well and differential rotation, by tidal disruption, or stellar winds
, the turbulence broadens the $^{12}$CO line and yields a larger velocity gradient. 
In combination with low density ($n_{\textrm{H}_2} \leq 10^3$ cm$^{-3}$), this reduces the optical depth considerably and the diffuse gas is difficult to detect in $^{13}$CO \citep[][]{Huttemeister2000,Sakamoto2007,Costagliola2011}.
%
%
%

Apart from opacity considerations, but related to the diffuse gas component, beam filling influences the line ratio as well. A large line ratio can be explained by $^{13}$CO tracing the dense cloud core, which is surrounded by a less dense / diffuse envelope only detectable in $^{12}$CO \citep{Sakamoto2007}.

All in all, high $R_{10}$ and $R_{21}$ line ratios can be explained by a combination of emission from a warm (and moderately dense) and a diffuse gas component
\citep{Aalto1995,Paglione2001,Aalto2007,Bolatto2013}.  
This is very likely the case in galactic nuclei where the GMCs are expected to be warmer and more turbulent \citep{Huttemeister2000,Paglione2001,Costagliola2011}. 

For our two galaxies with $^{13}$CO(2--1) observation, we can interpret the conditions of the ISM as follows.
In both galaxies, the peak in $^{13}$CO(2--1) emission is single-sided and slightly off-center from the $^{12}$CO(2--1) peak. Their locations seem to be in the vicinity of the overdensity zone where the spiral arm/dust lane in the bar connects to the circumnuclear region. Therefore, their presence could be related to a pile-up of gas near the intersection of the $x_1$ and $x_2$ orbits: the densities along the shock front of the bar spiral arms increase toward the center. Their maxima near the $x_1$-$x_2$ intersection form a bar-like, double-peaked structure that is perpendicular to the primary bar and observable in $^{12}$CO \citep{Kenney1992,Kohno1999} and higher density tracers \citep[see cartoons in][]{Meier2005,Meier2012}.
So there seems to be a sufficiently large region of nondiffuse, moderately dense gas in the vicinity of the $x_1$-$x_2$ orbit-crowding regions so that the $^{13}$CO emission becomes detectable on our sensitivity level and beam size (beam filling). The asymmetry in the emission might be due to a difference in excitation or filling factor between the two arms \citep[see cartoons in][]{Meier2005}. 
The $^{12}$CO/$^{13}$CO(1--0) ratio in HE 0433-1028 reaches a Galactic value of $R_{21} \sim 7$ in the $^{13}$CO peak and $R_{21} \sim 12$ in the region opposite of the $^{12}$CO peak, suggesting more turbulent (larger region of enhanced velocity dispersion, see map), diffuse, or warmer gas.

The central depression in $^{13}$CO and the corresponding ratio of $R_{21} \sim 20$ is most likely attributed to a significantly higher temperature within the circumnuclear region, leaving the lower $^{13}$CO transitions rather unpopulated.
In the case of HE 1108-2813, the $^{12}$CO/$^{13}$CO(1--0) ratio is already high in the $^{13}$CO peak with $R_{21} \sim 20$ and reaches even higher values at the center ($R_{21} \sim 25$) and opposite the center ($R_{21} \gtrsim 30$), indicating a generally high amount of diffuse gas in the circumnuclear region and beyond, with a 
higher temperature in the very center. 

Toward the edges of the bar in both galaxies the ratios decrease toward quiescent Galactic values ($R_{21} \sim 5$). The $^{13}$CO emission in both galaxies is close to the detection limit, therefore, the features found need to be verified. 

In addition, the emission spots in the northern spiral arm of HE 1108-2813 need to be tested. They are GMC-like ($R_{21} \sim 5$) and located downstream of the dust lane shock, a typical location for dense gas, compressed by the spiral arm passage, and for consequential star formation sites. 
The stellar associations found in that region, however, (see Fig. \ref{HE1108-overlay}) are very small compared to other sites along the bar where no $^{13}$CO was detected. Hence, the reliability of that northern arm $^{13}$CO emission is highly speculative.

\subsection{$^{12}$CO(3--2)/HCO$^+$(4--3)}

Studies on this specific line ratio, i.e., $^{12}$CO(3--2)/HCO$^+$(4-3), are rare in the literature. For example, \citet{Wilson2008} find an average ratio 19 $\pm$ 9 for their (U)LIRG sample. Our lower limit $^{12}$CO(3--2)/HCO$^+$(4--3) ratio > 16 suggests that our SNR is insufficient to detect HCO$^+$(4--3), therefore, we cannot prove the presence of the very dense gas ($n_\textrm{crit} \sim$ 6.5 $\times$ 10$^6$ cm$^{-3}$) with our observations.

\subsection{$^{12}$CO $J+1 \leq 3$ ratios}
\label{sec:12COratios}

A $^{12}$CO(3--2)/(1--0) ratio of $r_{31} \sim 1$ is common in the inner kiloparsecs of galaxies with enhanced central star formation and for starburst galaxies \citep{Devereux1994,Dumke2001,Muraoka2007,Papadopoulos2008,Iono2009}, 
whereas the average values of $r_{31}$ in nearby galaxies are on the order of 0.6 \citep{Mauersberger1999,Israel2005}. The $^{12}$CO(3--2) emission intensifies toward star-forming regions and the $^{12}$CO(3--2)/(1--0) ratio increases from the outskirts of the galaxy toward the center \citep{Dumke2001}. 


HE 1029-1831 shows line ratios of $r_{32} \sim 2$, $r_{21} \sim 0.5$, implying the onset of a higher excitation phase at the $J+1=3$ level with a well-excited and optically thin global $^{12}$CO spectral line energy distribution \citep[SLED; generally, $r_{32}\geq r_{21}$ and $r_{21}\leq 0.6 - 1 $;][]{Papadopoulos2012a}.
%
%
In order to thermalize the $^{12}$CO(3--2) line, a density of $n_\textrm{crit}$ $\sim$ few $10^4$ cm$^{-3}$ is needed. Keeping its emission optically thin requires a kinetic temperature of $T_\textrm{k} \geq 100 $ K and highly turbulent gas motion.

A subthermal low-excitation phase $^{12}$CO(2--1)/(1--0) ratio $r_{21} < 0.6,$ and potentially $r_{32} < 0.3$ , is ambiguous concerning the ISM state. This phase can either imply a cold ($T_\textrm{k} \leq 20$ K) and virialized 
gas state, as known for quiescent Galactic GMCs, or can hint at a warm ($T_\textrm{k} \gtrsim 30$ K) and nonself-gravitating state. 

This matches well our findings. All three galaxies have a $r_{21} \lesssim 0.6$. For two of these galaxies, the presence of a large fraction of diffuse gas in the circumnuclear region is indicated and the third exhibits a high-excitation phase by $r_{32} \sim 2$ overtaking the ambiguous low-excitation phase state ($r_{21} \sim 0.5$). This is not unexpected because we find starburst features in all three galaxies (see Sect. \ref{sec:SF}) and, for HE 1029-1831, the starburst has been confirmed by NIR data \citep{Busch2014}. 

Furthermore, their FIR luminosities $L_\textrm{FIR}$ are very similar. 
Assuming that the starburst is vigorous enough to dominate the FIR, the tight correlation of $L_\textrm{FIR}$, and $L_\textrm{CO}$, and $L_\textrm{CO(3--2)}$, in particular \citep[tracing warmer denser, potentially star-forming gas;][]{Iono2009,Wilson2012}, suggests the two $^{13}$CO-tested galaxies show a $^{12}$CO(3--2)/(2--1) ratio $r_{32} \gtrsim 1$ in the circumnuclear region similar to HE 1029-1831. In addition, we can expect the $R_{21}$ in HE 1029-1831 to show a large value characteristic for LIRG/SB.

\citet{Papadopoulos2012b,Papadopoulos2012a} identify AGN, cosmic rays (CR), and turbulence as the only viable large scale heating sources in (U)LIRGs. In contrast to the spatially confined star formation sites, these energy sources are efficient enough to raise the temperature and density of an entire galaxy-sized molecular gas reservoir so that the high-excitation $^{12}$CO SLED, seen in starburst galaxies/ULIRGs, emerge.

Though our three galaxies might not be extreme in starburst, we expect CR and turbulence to play an important role, since these mechanisms seem to be already dominating the heating of the molecular gas in the Galactic center \citep[e.g.,][]{YZ2007,Goicoechea2013}, i.e., the center of a normal galaxy.

\subsection{$\alpha_\textrm{CO}$ - factor and diffuse gas}
\label{sec:Xco}

The mass is related to the $^{12}$CO luminosity by the conversion factor $\alpha_\textrm{CO}$. 
The basic idea behind this factor is to "count clouds" in the telescope beam under the assumption that the GMCs are virialized
\citep[optically thin ensemble;][]{Bolatto2013}.
%
%

However, in the centers of local galaxies $\alpha_\textrm{CO}$ is on average lower by a factor of 2 than in the disk and the central $\alpha_\textrm{CO}$ of some of these galaxies are 5-10 times lower than in the center of the Milky Way \citep[][]{Sandstrom2013}.
The conversion factor correlates with the SFR surface density and the molecular gas density \citep[][]{Casey2014},
and anticorrelates with the $^{12}$CO/$^{13}$CO ratios that increase toward the centers of galaxies \citep{Paglione2001}. Therefore, a reduced $\alpha_\textrm{CO}$ value seems to have the same origin as a high $^{12}$CO/$^{13}$CO ratio R discussed in Section \ref{sec:1213CO}, i.e., diffuse gas.
%
%

Evidence has been found in recent years that $\gtrsim$ 30 $\%$ of the $^{12}$CO luminosity emerges from diffuse gas, or at least gas with densities $\lesssim 10^4$ cm$^{-3}$ \citep[][]{Solomon1989,Wilson1994,Rosolowsky2007,Hughes2013,Schinnerer2013,Pety2013,Shetty2014}, and seems to be similar in extent and velocity dispersion to the \ion{H}{i} disk, making up a single dynamical component \citep[][]{Dame1994,Caldu2013}.
%
%

Hence, the basic assumptions made for the standard conversion factor do not hold anymore in regions with high SFR surface density, i.e., centers of galaxies and regions of vigorous star formation. For example, for the highly turbulent ISM in (U)LIRGs, a standard $\alpha_\textrm{CO}$ 
of 4.3 - 4.8 $M_\odot$~(K km s$^{-1}$ pc$^2$)$^{-1}$ \citep[][respectively]{Bolatto2013,Solomon1991} 
assuming entirely self-gravitating gas overestimates the gas mass by a factor of $\sim$ 6 \citep[][]{Downes1998, Papadopoulos2012b}.

However, the conversion factor for ULIRGs based on low-J $^{12}$CO emission can severely underestimate the gas mass, when a large fraction of the gas resides at densities that are too high \citep[i.e., $n_{H_2}$ $\geq$ 10$^4$ cm$^{-3}$;][]{Solomon1992,Gao2004} to significantly contribute to the low-J $^{12}$CO emission, or when an existing large reservoir of cold, star formation quiescent gas is so faint in low-J $^{12}$CO emission that the global $^{12}$CO SLED is dominated by the central starburst \citep{Papadopoulos2012b}.

Spatially resolved low-J $^{12}$CO and $^{13}$CO lines and dust continuum emission help to track the ISM properties and the $\alpha_\textrm{CO}$ gradient also in order to separate the cold gas from the central star-forming gas component. 
In our low-resolution data of HE 0433-1028 and HE 1108-2813, the star formation region and the cold disk are vaguely indicated by the $^{12}$CO/$^{13}$CO line ratio distribution. 
Thus, we can conclude that the center is described by a rather low, starburst/ULIRG-like $\alpha_\textrm{CO}$, whereas the material in the primary bar is characterized by a higher, Milky Way (MW) like $\alpha_\textrm{CO}$.
Assuming 20\% of the $^{12}$CO luminosity resides in the primary bar with $\alpha_\textrm{CO MW} =$ 4.8 $M_\odot$~(K km s$^{-1}$ pc$^2$)$^{-1}$ and 80\% in the central region (results of the Gaussian component for HE 1108-2813; Table \ref{CO-source-fit}) with $\alpha_\textrm{CO ULIRG} =$ 0.8 $M_\odot$~(K km s$^{-1}$ pc$^2$)$^{-1}$, we obtain a global value of $\alpha_\textrm{CO} =$ 2 $\alpha_\textrm{CO ULIRG} =$ 1.6 $M_\odot$~(K km s$^{-1}$ pc$^2$)$^{-1}$.

Aside from the $^{12}$CO/$^{13}$CO ratio, the importance of diffuse gas in our observed galaxies is also implied by the lower FWZI compared to the single dish observations, which is consistent with results from \citet{Caldu2015}, where the interferometric line widths are 20 - 40 \% less than the single dish widths. 
This can either hint at diffuse gas with a high velocity dispersion, which seems to be similar to that of the \ion{H}{i} disk \citep[][]{Caldu2013}, or at large regular structure of small molecular clouds separated by less than a beam size with the high dispersion representing the cloud-cloud motion. In both cases, the corresponding emission would be resolved out, when the overall extent is larger than the maximum angular scale detectable by the interferometer.

\section{Summary and conclusions}
\label{sec:sum}

We have observed three galaxies from the LLQSO sample with the SMA in Hawaii with a resolution of 1.5$''$ - 4.5$''$. HE 0433-1028 and HE 1108-2813 were observed in $^{12}$CO(2--1) and $^{13}$CO(2--1), and HE 1029-1831 in $^{12}$CO(3--2) and HCO$^+$(4--3).
In the following we summarize the results and conclusions: 

\begin{itemize}

\item \textit{Morphology:} In all three galaxies, the bulk of the detected molecular gas, i.e., > 80\%, is confined to a region with < 1.8 kpc in diameter (i.e., the FWHM of the Gaussian component fit to the central bulk of emission). The gas partly extends along the primary bar and is not detected beyond the bar edges at our sensitivity level. In HE 1108-2813, the $^{12}$CO(2--1) emission traces the full bar including the spiral arms within it. The central emission bulk appears in all three galaxies to be extended toward the incoming dust lanes at the leading edges of the bar, implying an accumulation of inflowing gas and dust near or at the connection points, i.e., the $x_1$ and $x_2$ orbit intersections. This is supported by the tentative detection of $^{13}$CO in these locations. HCO$^+$(4--3) has not been detected in HE 1029-1831.

\item \textit{Kinematics:} Along the gradient, the velocities show a steep slope with indications for an unresolved central component reaching velocities $\geq$ 100 km s$^{-1}$ , which could be the same feature seen perpendicular to the velocity gradient. In HE 1108-2813 only, the sensitivity is sufficient to outline the rotation curve to the bar tips. The prominent x-shape is evident for a kinematically decoupled component in the center, such as a nuclear disk.

\item \textit{Masses:} We obtain dynamical masses of $M_{\textrm{dyn}} = (1.5 - 6.7) \times 10^9 M_\odot$. 
The mass range of the detected molecular gas is $M_\textrm{mol} = (0.7 - 6.2) \times 10^9 M_\odot$ and corresponds to $10 - 50~\%$ $M_{\textrm{dyn}}$, for ULIRG mass conversion factor, and $80 - 300~\%$ $M_{\textrm{dyn}}$, for Milky Way mass conversion factor. However, the significance of $M_{\textrm{dyn}}$ is diluted by a low SNR and low accuracy of the inclination. For HE 1108-2813 only, we obtain $M_\textrm{mol} < M_{\textrm{dyn}}$, which is the least criterion that should be fulfilled.
The dust mass is constrained by an upper limit of $M_{\textrm{dust}}$ = (1.6 - 4.4) $\times$ $10^6$ $M_{\sun}$. This yields a gas-to-dust ratio of $M_{\textrm{mol}}$/$M_{\textrm{dust}}$ = 180 - 350 (ULIRG conversion factor) and 1100 - 3400 (Milky Way conversion factor). 

\item \textit{Star formation:} We derive star formation rates of $SFR=$ 14 - 33 $M_\odot$~yr$^{-1}$, resulting in upper limit consumption timescales of $\tau_\textrm{SFR} =$ 70 - 490 Myr. Together with minimum gas mass surface densities of $\Sigma_{mol}$ = 50 - 550 $M_\odot$~pc$^{-2}$ and SFR surface densities of $\Sigma_\textrm{SFR}$ = 1.1 - 3.1 $M_\odot$~kpc$^{-2}$yr$^{-1}$, the three galaxies can be expected to harbor a circumnuclear starburst.

\item \textit{Excitation:} $^{13}$CO is found next to the $x_1$ and $x_2$ intersection overdensities as expected. Brightness asymmetries can be caused by excitation differences and/or beam filling. The $^{12}$CO/$^{13}$CO(2--1)-ratio in the $^{13}$CO peak of HE 0433-1028 is $R_{21} \sim 7,$ which is typical for the Milky Way. The ratio
rises to $R_{21} \sim 20$ at the nucleus as a result of higher temperatures and stronger turbulence, which is often seen toward starbursts. 
The trend in the line ratio is the same for HE 1108-2813, but starts on a higher level with $R_{21} \sim 20$ at the $^{13}$CO peak and $R_{21} \sim 25$ in the nucleus. The gas is even more diffuse/hot elsewhere ($R_{21} \gtrsim 30$). Toward the bar edges the ratio reduces to Galactic values. HCO$^+$(4--3) was not detected, most likely due to a lack of sensitivity. 

The $^{12}$CO(3--2)/(1--0)-ratio of $r_{31} \sim 1$ seen in HE 1029-1831 is typical for enhanced central star formation in the inner kiloparsecs of galaxies and for starburst galaxies.
As a striking result, the decreasing sequence of line ratios, expected for a single ISM phase, reverses, i.e., $r_{32} \sim 2$, $r_{21} \sim 0.5$. Obviously, a second higher excited phase is present apart from the cold or diffuse gas phase implied by $r_{21}$.

Putting the information from all three galaxies together the following picture emerges. All three galaxies have a low $r_{21} \lesssim 0.65, $ which could be either attributed to cold virialized or diffuse nonvirialized gas. In two galaxies, we found evidence for a diffuse gas phase interpretation. 
Since all three galaxies are strong and similar in FIR emission, we can expect their $L_\textrm{CO(3--2)}$ to be similar ($L_\textrm{FIR}$ -- $L_\textrm{CO(3--2)}$ - correlation) as well as their central $r_{31}$ and $r_{32}$ (because of similar $L_\textrm{CO(1--0)}$). Conversely, we expect HE 1029-1831 rather to show high $R_{21}$ and $R_{10}$, i.e., the presence of a diffuse gas state.

As long as there is not more information on the gas excitation, the mass-to-luminosity ratio $\alpha_\textrm{CO}$ is highly vague.
Since it anticorrelates with the $^{12}$CO/$^{13}$CO(2--1)-ratio, our maps track spatial variations of $\alpha_\textrm{CO}$. Regions with high $R_{21}$ imply a low ULIRG-typical $\alpha_\textrm{CO}$. A large fraction of the gas is confined to these regions, therefore, the average $\alpha_\textrm{CO}$ for the galaxies might be in between ULIRG- and Milky Way-like values.
The dynamical and dust masses appear to support a low conversion factor. Nevertheless, there is the possibility that a potential large cold gas reservoir is simply outshone by the central starburst.
\end{itemize}

Our interferometric data allow us a first insight into the distribution of the molecular gas, its kinematics, and its excitation state in these three galaxies. 
This study demonstrates nicely the wealth of information on nearby (z $\sim$ 0.01 - 0.06) galaxies, which can already be obtained at medium spatial resolutions, i.e., 0.5 - 1 kpc-scales. These observations provide important parameters to be tracked in the scope of evolution studies, in addition to single dish data.
To constrain the ISM phases with higher accuracy, observations of higher CO transitions and high density tracers (HCN, HCO$^+$, etc.) at several transitions are mandatory. To follow the matter transport to the nucleus and the distribution of the ISM states, we need to map the galaxies at a far higher resolution. 
With the Atacama Large Millimeter/submillimeter Array (ALMA) we can access spatial scales around 100 pc and less (e.g. 0.1$''$ $\lesssim$ 80 pc for the three galaxies) so that we can directly compare our results to local samples, such as the 
NUGA sample observed both with the IRAM PdBI \citep{GB2003b,Combes2004,Hunt2008,Casasola2008,Haan2009,Casasola2010,Casasola2011c,GB2012} and with ALMA \citep[e.g.][]{Combes2013,Combes2014,GB2014}.
In addition, observations at comparable scales in the NIR complete the analysis with information on the ionized gas, hot molecular gas, stellar component, and supermassive black hole \citep[e.g.,][]{Smajic2014,Busch2014,Smajic2015}. Above all, these studies need to be extended to a larger fraction of our sample to yield representative results.

\hyphenation{For-schungs-gemein-schaft}

\begin{acknowledgements}
The authors thank the anonymous referee for constructive comments.
This work was supported in part by the Deutsche Forschungsgemeinschaft (DFG) via SFB 956. G. Busch is member of the Bonn-Cologne Graduate School of Physics and Astronomy (BCGS). L. Moser and S. Smaji\'c are members of the International Max Planck Research School (IMPRS) for Astronomy and Astrophysics Bonn/Cologne. J. Scharwächter acknowledges the European Research Council for the Advanced Grant Program Number 267399-Momentum. M. Valencia-S. received funding from the European Union Seventh Framework Programme (FP7/2007-2013) under grant agreement No. 312789.
The Submillimeter Array is a joint project between the Smithsonian Astrophysical Observatory and the Academia Sinica Institute of Astronomy and Astrophysics and is funded by the Smithsonian Institution and the Academia Sinica.
This research has made use of the NASA/IPAC Extragalactic Database (NED), which is operated by the Jet Propulsion Laboratory, California Institute of Technology, under contract with the National Aeronautics and Space Administration.
This research is also based on observations made with the NASA/ESA Hubble Space Telescope, obtained from the data archive at the Space Telescope Science Institute. STScI is operated by the Association of Universities for Research in Astronomy, Inc. under NASA contract NAS 5-26555.

\end{acknowledgements}


\bibliographystyle{aa} 
\bibliography{LMoser_QSOs}
\end{document}